# Infrared-terahertz double resonance spectroscopy of $CH_3F$ and $CH_3Cl$ at atmospheric pressure

Dane J. Phillips,[*] Elizabeth A. Tanner,[†] Frank C. De Lucia,[‡] and Henry O. Everitt[§]

[*] *Kratos – Digital Fusion, 4904 Research Drive, Huntsville, AL 35805, USA*

[†] *IERUS Technologies, 2904 Westcorp Blvd. Ste 210, Huntsville, AL 35805*

[‡] *Department of Physics, 191 Woodruff Ave., The Ohio State University, Columbus, OH 43210*

[§] *Charles M. Bowden Laboratory, Army Aviation and Missile RD&E Center, Redstone Arsenal, AL 35898*

**Abstract**

A new method for highly selective remote sensing of atmospheric trace polar molecular gases is described. Based on infrared/terahertz double resonance spectroscopic techniques, the molecule-specific coincidence between the lines of a $CO_2$ laser and rotational-vibrational molecular absorption transitions provide two dimensions of recognition specificity: infrared coincidence frequency and the corresponding terahertz frequency whose absorption strength is modulated by the laser. Atmospheric pressure broadening expands the molecular recognition "specificity matrix" by simultaneously relaxing the infrared coincidence requirement and strengthening the corresponding terahertz signature. Representative double resonance spectra are calculated for prototypical molecules $CH_3F$ and $CH_3Cl$ and their principal isotopomers, from which a heuristic model is developed to estimate the specificity matrix and double resonance signature strength for any polar molecule.

Receipt date: ???

PACS indexing codes: 33.20.-t, 33.40.+f

I. INTRODUCTION

Double resonance (DR) techniques are widely used in all phases of optical spectroscopy, from ultrafast pump/probe and excitation correlation measurements in the ultraviolet/visible/infrared spectral regions to investigations of molecular collisional physics in the microwave/microwave and infrared/millimeter wave regions [1 - 7]. Although wavelength-degenerate DR techniques provide some insight into the relaxation dynamics from a photoexcited state, wavelength-nondegenerate DR techniques - which rely on the ability to tune the frequencies of the pump and probe beams independently - provide greater insight into energy transfer pathways and rates [8 - 11]. Sometimes, the limited frequency tunability of pump or probe sources limits the effectiveness of DR techniques. For example, optically pumped far infrared lasers - which use a line-tunable $CO_2$ infrared (IR) laser to excite a terahertz (THz) rotational population inversion in a low pressure gas - are constrained by the rare coincidence (< 100 MHz) between any $CO_2$ laser line and any molecular rotational-vibrational IR transition [12 - 14]. This rare coincidence forces optically pumped far-IR (OPFIR) lasers to operate on only a few THz frequencies per molecule, but as the gas pressure and corresponding IR and THz linewidths increase, greater tunability and spectral coverage is possible [15 - 20].

Recently we have proposed that the IR/THz DR technique may be adapted for remote sensing of trace gases in the atmosphere if IR pulses of sufficiently short duration (~ 100 ps) can be generated [21]. In particular, it was shown that a 100 m thick cloud of $CH_3F$ with a uniform concentration of 1 ppm (~1 mTorr partial pressure) could be detected using a THz transceiver tuned to the frequency of the IR pump-induced THz DR signature. Detection occurs through the co-propagation of IR pump and THz probe beams through the trace gas cloud to a retro-reflecting surface up to 1 km away. If absorbed by the trace gas, the pulsed IR pump beam briefly alters the absorption strength of the returned continuous wave THz probe beam, and this repeating modulation may be detected at the pulse repetition frequency of the laser. The actual sensitivity of this technique depends in a complex manner upon the IR and THz absorption coefficients of the molecule and the atmospheric water vapor content. Recognition specificity is achieved through the molecule-unique combination of IR pump coincidence frequency and the frequency of the pump-induced change in the THz signature. For a given molecule, only a few of its hundreds of IR ro-vibrational transitions are coincident with any $CO_2$ laser line, and each rare coincidence produces a unique pump-induced absorption change in a few of the hundreds of THz rotational transition frequencies. These two dimensions of recognition (IR pump and THz DR frequencies) comprise a "specificity matrix" whose sparseness can be used to identify an atmospheric trace gas remotely with enough spectral discrimination that even isotopomers may be distinguished.

One of the findings of that study [21] is that the coincidence requirement between $CO_2$ laser and IR molecular transition is relaxed when the trace gas primarily collides with atmospheric pressure gases (see Figure 1). Specifically, Figure 2 indicates that the ro-vibrational IR linewidth of $CH_3F$ evolves from a Doppler-broadened $\Delta v_D$ = 33 MHz (half-width at half-maximum, or HWHM) for low pressure operation (typical of OPFIR lasers) to a pressure broadened $\Delta v_P \approx$ 3 MHz/Torr = 2.3 GHz HWHM when the same 1 mTorr of gas is in an atmosphere of $N_2$ and $O_2$ [22]. Moreover, the IR linewidth function changes from Gaussian to Lorentzian, meaning that the spectral overlap falls off less abruptly with frequency at atmospheric pressure. This broadening relaxes the pump coincidence requirement from those ro-vibrational lines within 66 MHz of a $CO_2$ laser line (spaced 24 to 69 GHz apart, see Figure 1a) to those within 4.6 GHz [23].  The benefit is that favorable IR/THz pump/probe combinations, including transitions that are not accessible at low pressure, may be selected for atmospheric remote sensing.

In order to explore the molecule-unique signatures that constitute the recognition specificity matrix, we calculate the double resonance signatures of all IR/THz pump/probe coincidences at atmospheric pressure for $^{12}CH_3F$ and $CH_3^{35}Cl$ and their principal isotopomers $^{13}CH_3F$ and $CH_3^{37}Cl$ that are 1% and 24% abundant, respectively.  These molecules were chosen as prototypical because their excited vibrational levels are well isolated but different in nature:  a $CO_2$ laser that is line tunable over 9-11 µm wavelength region (900-1100 cm$^{-1}$) excites the $V_3$ stretching mode of $CH_3F$ and the $V_6$ bending mode of $CH_3Cl$.  We will show that the DR signatures are dramatically different for these two types of vibrational states in ways that are unique to the molecule, providing additional recognition specificity. Because many new coincidences occur at atmospheric pressure, we show how to estimate the strength of the DR signals so that the IR/THz combinations most useful for remote sensing may be identified.

## II. DOUBLE RESONANCE SPECTROSCOPY

### A.  Pump Coincidences

The energy levels of the ground and excited vibrational states of $CH_3F$ and $CH_3Cl$ were calculated using the rotational constants listed in Tables I and II. For $CH_3F$, the standard P-type ($\Delta J = -1$), Q-type ($\Delta J = 0$), and R-type ($\Delta J = +1$) transitions from ground state to the $V_3$ vibrational level produce a classic stretching mode spectrum (Figure 1b) whose absorption strengths are determined by a combination of the ground state's fraction of the temperature-dependent partition function, quantum mechanical transition matrix element (Table III), and dipole derivative for the vibrational transition. The notation for a ro-vibrational transition is $X_K(J)$, where J,K designate the rotational state of origin in the ground vibrational level and X = P, Q, or R depending on the type of ro-vibrational transition excited.

By comparison, the spectrum of the $V_6$ bending mode is characterized by *l*-doubling that couples the bending motion $l = \pm 1$ with the rotation designated by the *K* quantum number. The notation is modified to $^YX_K(J)$ to account for whether the ro-vibrational transition excited in $V_6$ satisfies $l = +1$ (Y = R) or -1 (Y = P). The corresponding spectrum (Figure 1c) is remarkably different than for the bending mode in $CH_3F$ because *l*-doubling breaks the $\pm K$ degeneracy, dramatically affecting the energy level structure and transition matrix elements (Table III). An examination of a partial low pressure rotational spectrum of these two excited states (Figures 1d & 1e) shows the effects of *l*-doubling and hyperfine splitting in $CH_3Cl$. (Hyperfine splittings in $CH_3Cl$ that are resolved at low pressure may be ignored at the atmospheric pressures of interest here.)

Using these calculated energy levels, the allowed IR ro-vibrational transition frequencies were calculated and compared with the known frequencies produced by line-tunable $CO_2$ lasers [23]. To find nearly overlapping coincidences, assume for the moment that the IR pump is a continuous wave $CO_2$ laser so the coincidence occurs within the linewidth of the molecular ro-vibrational transition. For a pure, 1 mTorr gas of $CH_3F$ or $CH_3Cl$ at 300K, the salient IR ro-vibrational transition is Doppler broadened with a Gaussian HWHM linewidth $\Delta v_D$ of 33.4 MHz and 27.4 MHz, respectively [24]. All ro-vibrational IR frequencies are then calculated, and the "offset" frequencies $v_{IR} - v_{pump}$ of those nearly coincident with specific $CO_2$ laser lines are compared with the values known from extensive research in OPFIR lasers [25, 26]. The results, shown in Table IV, indicate the offset frequencies are good agreement when the vibrational energies in Table I are used. As a notation to simplify the discussion, the rotational states connected by the pump are labeled $(J_p, K_p, V_0)$ and $(J_p', K_p', V_i)$. The principal rotational transitions exhibiting THz DR signatures, involving the two pairs of adjacent rotational states $(J_p \pm 1, K_p, V_0)$ and $(J_p' \pm 1, K_p', V_i)$, are labeled $R_0^-$ and $R_0^+$, $R_i^-$ and $R_i^+$, respectively (see Figure 3). The frequencies of these four rotational transitions are presented in Table IV.

Next, the ro-vibrational transitions of $CH_3F$ or $CH_3Cl$ with 1 mTorr of partial pressure are broadened by an atmosphere of $N_2$ and $O_2$ (i.e. concentration $\theta \sim$ 1 ppm of analyte), and the coincidence calculation is repeated. Using a pressure broadening parameter of 3 MHz/Torr FWHM - typical for hard shell collisions with non-polar molecules [22] - the estimated linewidth of $\Delta v_p$ = 2.3 GHz HWHM far exceeds the Doppler-broadened linewidth. Using a Lorentzian lineshape function with this linewidth parameter, the overlap expands and many new coincidences are found.

Table V lists only those coincidences in $^{12}CH_3F$ and $^{13}CH_3F$ whose overlap with the $CO_2$ laser is within one atmospheric pressure-broadened Lorentzian linewidth of the ro-vibrational center frequency. Of the new ro-vibrational coincidences that arise, many involve $CO_2$ lines that experience no low pressure coincidences, while the rest represent additional coincidences with $CO_2$ lines already known to have a low pressure coincidence. In both cases, atmospheric pressure broadening of the ro-vibrational transitions often causes a given laser line to excite many rotational states simultaneously. The four rotational frequencies reported in Table V correspond to the ro-vibrational transition whose $(J_p, K_p, V_0) \rightarrow (J_p', K_p', V_i)$ transition most closely coincides with the strongest modulated terahertz absorption.

B. Differential Absorption

The strength of the pump-induced THz DR signatures may be calculated from the standard equation for molecular absorption [24, 27]

$$\alpha = \frac{8\pi^3}{3hc} \nu (n_L - n_U) |\mu|^2 S(\nu, \nu_0)$$
$$= 3.24 \times 10^{-6} \frac{\nu}{\Delta \nu} (f_L - f_U) \theta |\langle U | \mu_t | L \rangle|^2 \ cm^{-1} \quad , \quad (1)$$

where $\nu$ is the frequency (GHz) of the transition and $n_i = 2.45 \times 10^{13} f_i \theta \ cm^{-3}$ is the population of the molecular states connected by the probe and pump ($i = L, U$) for a non-degenerate fractional state population $f_i$ and concentration $\theta$ ppm at 300K in an ambient atmosphere of 760 Torr. The factor containing $\mu_t$ is the dipole matrix element for this transition (Debye), and $S(\nu, \nu_0)$ is the lineshape function that simplifies to $1/\pi\Delta\nu$ (GHz$^{-1}$) at the peak of a pressure-broadened Lorentzian lineshape [28, 29]. The absorption $\alpha_{THz}$ at $\nu_{THz}$ arises from rotational transitions mediated by the permanent dipole moment of the molecule ($\mu_t = \mu$), while the absorption $\alpha_{IR}$ at $\nu_{IR}$ arises from ro-vibrational transitions mediated by the appropriate dipole moment derivative along the respective normal coordinate ($Q_i$) for the corresponding excited vibrational level ($\mu_t = d\mu/dQ_i$) [28].

Note that the absorption coefficient is proportional to the square of the appropriate dipole moment mediating the transition, multiplied by the rotational transition branching ratio $0 < |\langle U|\mu|L \rangle|^2 < 1$ given in Table III [30]. For IR transitions, the dipole derivative can be a factor of 10 (stretching modes) to 100 (bending modes) times weaker than the dipole moment itself, an effect that will tremendously affect DR spectroscopy. Specifically, $\mu = 1.86$ D and $d\mu/dQ_i =$ 0.2756 D for the CH$_3$F V$_3$ transition, while $\mu$ = 1.90 D and $d\mu/dQ_6$ = -0.0388 D for the CH$_3$Cl V$_6$ transition [29 - 31]. Consequently, the IR transition absorption strength is approximately two orders of magnitude stronger in the stretching V$_3$ mode than in the bending V$_6$ mode. This difference arises because of the type of motion involved, not because of the differing masses of the two molecules, suggesting a more universal insight: DR signatures will generally be stronger in stretching modes than bending modes. Unfortunately, molecules tend to have more bending modes than stretching modes, which can be challenging if those modes span the region over which a CO$_2$ laser can be tuned.

The strength of the pump-induced THz DR signature may now be estimated. For an optically thin cloud ($\alpha_{THz}d \ll 1$) diluted in air to concentration $\theta$ with IR optical depth $\alpha_{IR}d \ll 1$, the total number of photons absorbed (i.e. the number of molecules photo-excited per pulse) is

$$N_{pumped} = \frac{\varepsilon}{h\nu_{IR}} \alpha_{IR} d \ , \qquad (2)$$

where $\varepsilon$ is the pump energy per pulse (J), $d$ is the path length of the laser in the cloud, and $\alpha_{IR}$ can be calculated from (1) with the assumption $n_U \approx 0$. From this, it is easy to show that the fraction of the $n_L$ molecules pumped per laser pulse in a beam of radius $r_{IR}$ (m) is

$$f_{pumped} = \frac{\varepsilon \alpha_{IR}}{(\pi r_{IR}^2)(h\nu_{IR})n_L} = 6.35 \frac{\left|\left\langle U \left| \frac{d\mu}{dQ} \right| L \right\rangle\right|^2}{\Delta \nu_{IR} r_{IR}^2} \varepsilon, \qquad (3)$$

where the transition matrix for this ro-vibrational transition is the product of the square of the corresponding dipole derivative (Debye) and the appropriate transition branching ratio Table III.

These calculations assume there is no pump saturation, but for highly intense pump beams saturation effects must be considered.

If the DR signature is observed on a time scale shorter than the collisional relaxation time (~100 ps at atmospheric pressure), then the only rotational transitions with photo-induced population changes are $R_0^+$, $R_0^-$, $R_i^+$, and $R_i^-$ (Figure 3). The pump-induced change in the absorption coefficients (1) arises from a change in the associated population differences from equilibrium ($n_L - n_U$) to pumped ($n_L \pm \Delta n - n_U$) values, where $\Delta n = N_{pumped}/V_{IR}$ and $V_{IR} = \pi r_{IR}^2 d$ is the volume occupied by the IR beam inside the gas. The measured THz DR spectrum is the pump-induced change in the rotational absorption coefficient

$$\Delta \alpha = \pm \frac{8\pi^2}{3h^2 c} \left[ \frac{\nu_{THz}}{\Delta \nu_{THz}} |\langle U | \mu | L \rangle|^2 \right] \left[ \frac{\Delta \nu_{IR}}{\Delta \nu_{IR} + (\nu_{IR} - \nu_{pump})^2} \frac{\alpha_{IR}}{V_{IR}(\pi r_{IR}^2)} \varepsilon \right] \quad (4)$$

derived from (1) and (2) as a product of terms that depend on the rotational transition (first bracket) and on the ro-vibrational transition (second bracket), where $\nu_{THz}$ corresponds to the monitored rotational transition with the indicated transition matrix element. Notice that a Lorentzian lineshape factor has been added to account for the reduced IR absorption caused by the offset (reported in Table V) between the pump and the atmospheric pressure broadened ro-vibrational transition frequencies. (Figure 2 indicates the pressure at which the Lorentzian lineshape must be replaced by a Gaussian.) Because $\Delta \alpha$ grows linearly with IR pump energy ($\varepsilon$) per pulse, the DR absorption strengths are reported as $\Delta\alpha/\varepsilon$ in units of kJ$^{-1}$ km$^{-1}$ in Table V.

### C. Overlap Enhancement

Since the THz rotational transitions undergo the same pressure broadening as the IR ro-vibrational transitions, the THz DR spectra of nearby pumped rotational transitions may also overlap. This overlap may strengthen or weaken the THz DR signal from the low pressure value (Eq. (4)) in which only one ro-vibrational transition is pumped. The multiplicative factor by which a given transition is strengthened or weakened is significantly less than the number of overlapping transitions because the constituent line strengths may have differing degeneracies and thermal populations, as well as overlaps with the opposite sign of $\Delta\alpha_j$. (For a given laser line, the overlapping transitions in $CH_3F$ are labeled by the subscripts in the "IR Transition" column of Tables IV and V.) The overlap enhancement factor in Table V

$$\Pi(\nu_j) = \frac{\sum_{i=1}^{n} \Delta\alpha_i \frac{(\Delta\nu_p)^2}{(\nu_i - \nu_j)^2 + (\Delta\nu_p)^2}}{\Delta\alpha_j}, \qquad (5)$$

estimates how many times stronger or weaker the DR signal $\Delta\alpha_j$ will be at frequency $\nu_j$ because of the overlap from nearby DR features of strength $\Delta\alpha_i$ located at $\nu_i$. Here, the summation is over all $n$ DR signatures associated with all IR transitions excited by a given laser line, where the $\Delta\alpha_i$ and $\Delta\alpha_j$ are calculated in (4). When the THz transitions overlap, the largest $\Delta\alpha_j$ necessarily produces the smallest value of $\Pi(\nu)$ at the transition labeled {R}, so this is the value provided in Table V. The peak of the summed DR signature will be near $\nu_{\{R\}}$, so this is the most appropriate frequency to use for the THz probe.

The product $\Delta\alpha_j \Pi(\nu_j)/\varepsilon$ represents the final estimate of the THz DR strength for a given IR/THz combination in the specificity matrix. This product can be used to ascertain which of the old or new coincidences are most promising for remote sensing for a given isotopomer.

Remember that $\Delta\alpha_j \Pi(\nu_j)$, which may be positive or negative, represents a pump energy-dependent ($\varepsilon$) increase or decrease in the ambient THz absorption strength as $\alpha_{THz}(\nu_j) \rightarrow \alpha_{THz}(\nu_j) + \Delta\alpha_j \Pi(\nu_j)$. Although usually observed as a short-lived pump-induced change in absorption strength, THz emission can actually occur when $-\Delta\alpha_j \Pi(\nu_j) > \alpha_{THz}(\nu_j)$. Because atmospheric collisional redistribution of population is minimal on timescales shorter than 100 ps, $\Delta\alpha_j \Pi(\nu_j)/\varepsilon$ only depends on the absorption strength of the IR ro-vibrational molecular transition, the detuning of the spectral overlap with the nearest $CO_2$ laser line, and the aggregate strength of the overlapping rotational transitions involving states simultaneously excited by the laser ($\Pi(\nu_j)$). Note that the absorption strength of the IR ro-vibrational transition depends on the type of vibrational mode for a given molecule (dipole derivative for stretching or bending modes), the specific ro-vibrational transition type involved (branching ratios for P, Q, or R), and the rotational partition function (or more precisely, the ground rotational state population $n_L$ of the pumped ro-vibrational transition). Not surprisingly, the optimal ro-vibrational transitions for DR spectroscopy are those with the strongest IR absorption, *i.e.* those nearest the peak of the IR spectrum.

Finally, it has already been pointed out that the pressure-broadened Lorentzian lineshape function falls off less abruptly than the Doppler-broadened Gaussian lineshape function, so overlaps are more readily excited. The decision to include only those overlaps within one linewidth $\Delta\nu_{IR}$ of the ro-vibrational line center was arbitrary. Recall the calculation assumed the $CO_2$ laser was operating continuously; however, if the laser is pulsed, the increased spectral bandwidth of the laser pulses should also be considered in the coincidence calculation. For the remote sensing application, pump pulses of duration comparable to the hard shell gas kinetic collision time (~100 ps) are required, corresponding to a spectral bandwidth of 10 GHz that is (not coincidentally) comparable to the pressure-broadened linewidth $\Delta \nu_P$ of the IR ro-vibrational transition. Clearly even more coincidences may be excited when such short pulses are used, but the pump efficiency is reduced by the fraction of pump radiation that falls outside the spectral bandwidth of the IR absorption line.

## III. SPECIFIC EXAMPLES

Having outlined the basic principles underlying the excitation of a DR signal, we next explore the variety of DR signal types, their strengths, and how they depend on the type of photo-excited ro-vibrational transition (Figure 3). Again, we will restrict ourselves to photoexcitation on timescales short compared to the fastest collisional relaxation time of the molecule (~100 ps), so only rotational transitions $R_0^-$, $R_0^+$, $R_i^-$, and $R_i^+$ are involved.

### A. Methyl Fluoride

We start with the simpler case, $CH_3F$, for which the stretching mode $V_3$ of both $^{12}CH_3F$ and $^{13}CH_3F$ isotopomers is photo-excited by a $CO_2$ laser. Considering first $^{12}CH_3F$, the 9P(20) line of a $CO_2$ laser is known to photo-excite the $Q_2(12)$ (and perhaps the $Q_1(12)$) transition at low pressure. The resulting DR signature has components at 612.409 ($R_0^-$), 663.365 ($R_0^+$), 604.297 ($R_3^-$), and 654.582 GHz ($R_3^+$). The absorption decreases ($\Delta\alpha_j < 0$) in transitions $R_0^+$ and $R_3^-$ and increases ($\Delta\alpha_j > 0$) in transitions $R_0^-$ and $R_3^+$ (Figures 3, 4a), illustrating that a characteristic Q-branch THz DR signature consists of four spectral features in the form of conjugate doublets (two $R^-$ and two $R^+$) separated by *~2B ≈ 51 GHz*. Assuming there is no IR pump saturation, equation (4) indicates the strengths of these four DR features are linearly proportional to the pump pulse intensity, their THz frequency, and the associated branching ratios (Table III). Each doublet consists of the identical rotational transitions in $V_0$ and $V_3$ separated in frequency by $\sim 2(B_3-B_0)(J + 1)$, a separation of 8.1 GHz for J = 11→12 and 8.8 GHz for J = 12→13. Given that $\Delta v_P$ = 2.3 GHz, these four features are well resolved at atmospheric pressure.

As noted above, the atmospheric pressure broadened width of the ro-vibrational transitions allows a given laser line to photoexcite many neighboring transitions, primarily those transitions with differing $K$ values, further increasing the strength of the DR signature. In $^{12}CH_3F$, the 9P(20) laser line now coincides with nine additional atmospheric pressure broadened ro-vibrational transitions ($Q_1(12) - Q_7(12)$, $Q_{13}(13)$, and $P_0(1)$ ). When a series of transitions with constant $J$ but varying $K$ are pumped (e.g. $Q_1(12) - Q_7(12)$), it is listed as a single entry in Table V with {R} = {K} for the transition with smallest $\Pi(v_j)$. Figure 2 shows how the multiplier $\Pi(v_K)$ for each of the $Q_K(12)$ transitions (K = 1-7) grows with increasing pressure, and Figure 4a and 5a show how their aggregate DR signature is the sum of all the constituent, overlapping Q-branch signatures for a portion of the THz region. Notice that $\Pi(v_j)$ = 1 for $Q_2(12)$ until the pressure broadening exceeds the Doppler broadening of the IR transition, after which there is a rapid, often non-monotonic growth in $\Pi(v_j)$ for all transitions as pressure increases. Notice $\Pi(v_j)$ for {K} = 6 only reaches 3.04 at atmospheric pressure, not the theoretical maximum of 6.15 if all seven K levels were equally populated and equally pumped. This reduction occurs because of two effects: the unfavorable branching ratios for low-K Q-branch transitions and the greater frequency separation of the rovibrational transition in the IR than the associated rotational transitions in the THz. For these same reasons, all other transitions have larger $\Pi(v_j)$ values because their isolated DR signature $\Delta\alpha_j$ is weaker. In other words, at atmospheric pressure the ro-vibrational transition that makes the largest contribution to $\Delta\alpha$ is $Q_6(12)$, not $Q_2(12)$ as at low pressure (Figure 4a).

At atmospheric pressure the 9P(20) line also excites other ro-vibrational transitions besides the $Q_K(12)$ series, namely $Q_{13}(13)$ and $P_0(1)$ (Figure 5a). These are quite weak because of the low thermal population in the respective $(J_p, K_p, V_0)$ states. Laser lines adjacent to 9P(20) also produce new coincidences, but the DR signatures are weaker (Figure 5a). This suggests a trend that will become more obvious when $CH_3Cl$ is studied: the overlap of ro-vibrational transitions and $CO_2$ laser lines can be likened to a Moire pattern formed by two overlapping grids with slightly varying spatial frequencies. Overlaps more often come in clusters of $CO_2$ laser lines instead of through random, isolated lines.

P- and R-type ro-vibrational transitions also manifest four THz DR components, but these occur as a triplet composed of a central doublet flanked by singlets of the opposite sign (Figure 3). The differing shapes of the P, Q, and R-dependent DR spectra, particularly the associated $\Delta\alpha_j$ sign differences, represent an additional degree of recognition specificity at a specific THz probe frequency. A typical example of a R-branch DR signature comes from the $R_3(4)$ transition in $^{13}CH_3F$, optically pumped by the 9P(32) line of a $CO_2$ laser (Figure 5b). A P-branch DR signature would be identical except flipped in sign. The central doublet consists of two components of the same sign separated by $\nu \sim 2(B_3-B_0)(J + 1) = 3.4$ GHz, while the singlets of opposite sign are separated from the doublets by $\nu \sim 2B \approx 50$ GHz. Consequently, the DR signature is spread twice as far in frequency as Q-branch DR signatures. The central doublet is barely resolved and very strong at atmospheric pressure.

It is striking how sparse the THz DR spectra are in Figures 5a and b. Although $^{12}CH_3F$ and $^{13}CH_3F$ have 13 and 22 pressure-broadened coincidences, respectively, most of them occur in clusters of adjacent ro-vibrational transitions of constant J but varying K excited by the same laser line. Consequently, there are only 5 and 2 distinguishable coincidences for $^{12}CH_3F$ and $^{13}CH_3F$, respectively, each of which has up to four frequency-resolvable THz DR components. No $CO_2$ laser line that excites one isotopomer excites the other. There are no DR signatures of one isotopomer within the same $2\Delta\nu_P \approx 5$ GHz window of any DR signature from the other isotopomer. Indeed, the recognition specificity matrices of $^{12}CH_3F$ and $^{13}CH_3F$ are so sparse that the isotopomers are easily distinguished through the selection of $CO_2$ laser line and corresponding THz probe frequency.

### B. Generalizations to Other Molecules

To explore how general a result this is for stretching mode DR spectra, consider more massive or complex prolate molecules and their correspondingly smaller rotational constants. Because individual absorption strengths $\alpha_{THz}$ will weaken with increasing molecular mass and rotational partition function, the correspondingly reduced $n_L$ and $\alpha_{IR}$ might also be expected to weaken the DR signals $\Delta\alpha_j \Pi(\nu_j)/\varepsilon$. However, the number of ro-vibrational IR coincidences and rotational state overlaps ($\Pi(\nu_j)$) increase with increasing molecular mass and complexity because the rotational spacings shrink and the peak THz absorption redshifts. These effects partially or completely compensate for the increased partition function and reduced $n_L$. The increasing number of coincidences with $CO_2$ laser lines will also add more elements to the recognition specificity matrix, permitting selection of favorable DR features in atmospheric transmission windows. Nevertheless, the specificity matrix is very large, containing approximately 60 x 200 x 2 = 24,000 unique combinations of sixty $CO_2$ laser frequencies, two hundred 5 GHz-wide windows from 0-1 THz, and +/- sign of $\Delta\alpha_j$, respectively. Only in the rarest of cases when two molecules excited by the same laser line and produce a DR signal of the same sign in the same 5 GHz-wide window will this affect the distinguishability of molecules through their DR signatures. Such false positive detections can be avoided by simply photo-exciting a different coincidence.

The challenge of multiple coincidences requires a strategy for identifying the strongest DR signature. Several insights from the analysis of $CH_3F$ arise. All isotopomers (and their isotopic abundance) must be considered when identifying the strongest DR signature for a given molecule. In the case of methyl fluoride, the largest $\Delta\alpha_j \Pi(v_j)/\varepsilon$ for the photo-excited R-branch ro-vibrational transitions of $^{13}CH_3F$ is 7 times larger than the strongest Q-branch-excited values in $^{12}CH_3F$. Only $^{13}C$'s rarity (1% isotopic abundance) prevents $^{13}CH_3F$ from being the preferred isotopomer.

Next, consider the factor $\left|\left\langle U | \frac{d\mu}{dQ} | L \right\rangle\right|^2$ in $\Delta\alpha_j \Pi(v_j)/\varepsilon$ for these ro-vibrational transitions, which may be expressed as the product of $(d\mu/dQ)^2$ and the quantum mechanical branching ratios in Table II. Because of the latter, Q-branch transitions are only favored when $J \approx K$. However, levels with $J \approx K$ have the weakest thermal rotational population, and these same branching ratios dramatically weaken the corresponding $\Delta J = 1$ DR signatures. For a given K, the branching ratio for Q-branch transitions tends to zero as J increases, while P- and R-branch transitions tend to ½. Because low *K* transitions have closer THz spectral spacing and comparatively larger thermal populations than high *K* transitions, it is evident that photo-exciting low *K* P- or R-branch transitions will yield the largest $\Pi_R$ enhancement factors.

Other factors also favor P- and R- branch transitions over Q-branch. It has already been mentioned that the strongest DR signatures will arise when transitions near the peaks of the IR spectra are excited. For increasingly massive molecules, the rotational quantum numbers $J$ associated with these peaks will increase, and the rotational constants $B$ and their differences ($B_i - B_0$) will decrease. Now, consider again the $2(B_3-B_0)(J + 1)$ spacings of the doublets in the respective transitions (ignoring higher order distortion terms) as illustrated in Figure 3. In general, features are resolvable when $2(B_3-B_0)(J + 1) > \Delta\nu_P$. In P- and R-branch transitions, the doublet components have the same sign of $\Delta\alpha$, so decreasing $(B_3-B_0)$ or $J$ simply increases the overlap and effectively doubles the strength of the central doublet. By contrast, the doublet components in Q-branch transitions have opposite sign, so decreasing $(B_3-B_0)$ or $J$ can weaken the doublet to the point of disappearance. This explains why Q-branch transitions often have DR signatures for which $\Pi(\nu_j)$ is less than one, and, more importantly, suggests that P- and R-branch transitions are more resilient than Q-branch transitions to the tradeoff between decreasing $(B_i - B_0)$ and increasing $J$.

Summarizing our findings to this point, the strongest DR signatures involving vibrational stretching modes will be obtained when photo-exciting rotational transitions with low $K$ whose $J$ values correspond to P- or R-branch ro-vibrational transitions near the peak of the corresponding IR spectra. Although these rules were derived for symmetric top molecules, their extension to the stretching modes of other types of molecules is straightforward and simplified by the fact that many of the asymmetric rotor distortions and hyperfine splittings are unresolved within the pressure broadened linewidth of the DR signal. These transitions constitute the "overlap enhancement factor" $\Pi(\nu_j)$ for asymmetric rotors.

## C. Methyl Chloride

Turning now to $CH_3Cl$, the calculated THz DR spectra in Figures 6 and 7 reveal that both isotopomers, $CH_3{}^{35}Cl$ and $CH_3{}^{37}Cl$, exhibit many more coincidences than in $CH_3F$. Although it is true that the $B$ rotational constant is roughly half that in $CH_3F$ (13.3 GHz for $CH_3Cl$, 25.5 GHz for $CH_3F$), that fact alone should only double the number of DR signatures over $CH_3F$. Instead, $CH_3{}^{35}Cl$ and $CH_3{}^{37}Cl$ have 154 and 140 coincidences within one atmospheric pressure broadened linewidth, respectively, 7-10 times more than $CH_3F$.

The difference is caused by the fact that the laser is now coincident with the $V_6$ vibrational bending level rather than the $V_3$ stretching level. Because bending modes couple vibrational ($l = \pm 1$) and rotational ($K$) motion to produce "$l$-doubled" spectra, the selection rules and branching ratios in Table III produce much richer ro-vibrational, rotational, and THz DR spectra. (Figure 1,6,7) In particular, the $\pm K$ degeneracy is removed by large energies that are proportional to $Kl$, and ro-vibrational transitions connect states of quantum number $K$ in $V_0$ with states of quantum number ($K$-$l$) in $V_6$. Besides doubling the number of rotational and ro-vibrational transitions, $l$-doubling introduces large energy and frequency splittings in $V_6$, so fewer ro-vibrational transitions occur within an atmospheric pressure broadened linewidth. Indeed, atmospheric pressure broadening never adds coincidences in $CH_3Cl$ with adjacent, ro-vibrational transitions that preserve $J$ but change $K$ for a given laser line, in stark contrast with $CH_3F$.

Instead, it is often found that a given laser line photo-excites a complex combination of overlapping and non-overlapping P-, Q-, and/or R-branch transitions simultaneously, typically involving widely separated rotational states and THz transition frequencies (see Figures 6 and 7).

The DR signature can be understood as a collection of these individual P-, Q-, and R-branch DR signatures (Figure 3) whose strengths and frequencies vary widely based on the idiosyncracies of the constituent rotational quantum numbers, energies, and transition branching ratios. The aggregate DR spectrum for a given laser line is quite unique, providing additional specificity.

For example, consider the 9P(26) laser line which has a known low pressure coincidence with the $^RQ_3(6)$ ro-vibrational transition in $CH_3^{35}Cl$ (Figure 4b). At atmospheric pressure, nine more P-, Q-, and R-branch ro-vibrational transitions are photo-excited, including the entire Q-branch sequence from $^RQ_3(4)$ to $^RQ_3(8)$, plus $^RR_2(7)$, $^RP_4(8)$, $^RR_0(24)$, $^PR_0(24)$, and $^RP_7(31)$, producing DR spectra that span 106 – 848 GHz. Interestingly, the strongest of these features are not those associated with the known low pressure $^RQ_3(6)$ coincidence but with the new $^RR_0(24)$ coincidence. A contributing reason for this is the increased overlap of the opposite sign Q-branch doublets spaced $2(B_i-B_0)(J + 1)$ apart: $(B_6-B_0) \approx 49$ MHz in $CH_3Cl$ while $(B_3-B_0) \approx 339$ MHz in $CH_3F$. Although the coincidences in $CH_3Cl$ typically involve states with larger *J* than $CH_3F$, atmospheric pressure broadening causes the Q-branch doublets to overlap strongly enough to reduce $\Pi(\nu_j)$ below 1 in most cases.

Comparing the relative strengths of the constituent transitions excited by the 9P(26) line provides helpful insights into the subtle interplay of factors that contribute to the complex DR signature. First of all, because the laser frequency is within one pressure-broadened linewidth of all ten coincident IR transitions, the strength of the corresponding DR signatures depends primarily on the branching ratio of the transition. Table III indicates that as $K \rightarrow J$, the IR transitions $^RR$ and $^PP$ grow much stronger than the $^PR$ and $^RP$ transitions. For example, the branching ratios of the IR transitions $^RR_2(7)$ and $^RP_4(8)$ are 0.229 and 0.022, respectively, explaining why their contributions to the DR signature differ by an order of magnitude (Figure 4b). Likewise, the $^RQ_K(J)$ branching ratios (Table III) tend to zero as $J \to K$, so inspection of the Q-branch series $^RQ_3(4)$ to $^RQ_3(8)$ in Figure 4b reveals the $V_6$ contributions to the Q-branch doublets disappear as $J$ decreases.

Figures 6 and 7 suggest the strongest DR feature for almost every coincidence derives from photoexcitation of a P- or R-branch transition. Although laser coincidences usually occur with isolated P- or R-branch transitions, sometimes a sequence of laser lines coincides with a sequence of IR transitions. A prominent example involves the series of laser lines spanning 10P(26) – 10P(38), where coincidences that preserve $K = 9$ for odd $J$ span $^PP_9(23)$ to $^PP_9(35)$, one per laser line. Many shorter sequences with this same Moire like overlap pattern of preserving $K$ but changing $J$ were found in both isotopomers.

By contrast, Q-branch transitions cluster strongly, so multiple IR transitions may be excited by a single laser line. Unlike $V_3$ of $CH_3F$ for which the P-, Q-, and R-branch IR transitions are arranged in clusters that preserve $J$ for all $K$ (Figure 1b), $l$-doubling in $V_6$ of $CH_3Cl$ spreads out the Q-branch transitions into clusters that preserve $K$ for all $J$ (Figure 1c). Consequently, a given $CO_2$ laser line may excite many such $J$-changing, $K$-preserving IR transitions, explaining how the 9P(26) line excites $^RQ_3(4)$ through $^RQ_3(8)$. When this occurs, an additional reduction in the DR signature occurs. Recall that the spacing of the transitions and the separation of the Q-branch components of opposite sign are both ~$2B$. As a result, the positive component of one DR feature overlaps the negative component of the preceding DR feature (Figure 4b), and the summed Q-branch DR signature is much weaker than its constituent components. Therefore, Q-branch transitions generally produce unfavorably weak DR features for bending vibrational modes, just as they did in stretching vibrational modes, but for different reasons.

So atmospheric pressure broadening typically produces overlapping sequences of Q-branch DR features with $\Pi(\nu_j) < 1$ and isolated P- and R-branch DR features with $1 < \Pi(\nu_j) < 2$. However, inspection of Table VI indicates that $\Pi(\nu_j)$ can sometimes be much larger. For example, $\Pi(\nu_j) \approx 25$ for the two DR signatures near 212 GHz associated with the $^RP_4(8)$ transition. Such $\Pi(\nu_j) > 2$ cases are not caused by the summation of overlapping DR signatures excited by a single IR transition, as was the case for $CH_3F$. Instead, these are rare instances in which very different ro-vibrational transitions excited by the same laser line produce THz DR signatures that coincidentally overlap. In the example at 212 GHz, the weak $^RP_4(8)$ DR signatures overlap with DR signatures 17 and 5 times stronger produced by the $^RR_2(7)$ and $^RQ_3(8)$ transitions, respectively.

The increased mass of $CH_3^{37}Cl$ lowers the vibrational energy of $V_6$, so the DR spectra of $CH_3^{37}Cl$ are similar in form but different in detail to those of $CH_3^{35}Cl$. The isotopic abundances (76% and 24% for $^{35}Cl$ and $^{37}Cl$, respectively) suggest that $CH_3^{35}Cl$ will be the preferred isotopomer for remote sensing. However, if a DR signature from $CH_3^{37}Cl$ is three times stronger than any from $CH_3^{35}Cl$, the less abundant isotope could produce the strongest DR signature. Indeed, if we constrain ourselves to look at the region of greatest practical interest below 300 GHz and account for isotopic abundances, we find that the strongest DR signature for $CH_3Cl$ is the $^RR_3(10)$ transition of $CH_3^{37}Cl$ at 287 GHz excited by the 9P(16) laser line.

## IV. CONCLUSIONS

Our analysis has shown the potential of IR/THz DR spectroscopy for the remote recognition of a trace gas in the atmosphere given sufficient pump intensity and gas concentration.  Discrimination comes from the two dimensions of a sparse recognition specificity matrix derived from (1) the few concidences that occur between $CO_2$ pump laser lines and ro-vibrational transitions of the analyte and (2) the unique changes in THz absorption the pump coincidences induce.  The strength of the DR signature depends sensitively on the type of ro-vibrational transition excited by the laser coincidence and the quantum and statistical mechanics of the involved transitions.  Atmospheric pressure broadening, which smears out transitions detected with traditional THz spectroscopic techniques, actually assists the IR/THz DR technique by providing more pump coincidences and more overlapping transitions for a given coincidence.  These advantages allow an optimal DR signature to be selected for a given analyte, and it was shown how to calculate this from basic molecular parameters.

Generally speaking, the DR signatures for $CH_3Cl$ are much weaker than those for $CH_3F$: the transition dipole moments are smaller and the rotational partition function is larger, so the ro-vibrational pump absorption coefficient is weaker.  Weak IR absorption, particularly for molecules whose pump coincidences involve bending modes, presents the greatest challenge to DR spectroscopy, just as it has limited the performance of many optically pumped far infrared laser gases.  For heavier molecules, this problem may be alleviated as the transition dipole moments tend to increase (e.g. $d\mu/dQ_6 = 0.054D$ for $CH_3Br$) and increasing complex vibrational energy level structures allow some selection in which vibrational mode the laser excites (e.g. a $CO_2$ laser may excite the $V_4$, $V_7$, or $3V^1_8$ vibrational modes of $CH_3CN$) [32, 33].

ACKNOWLEDGEMENTS

The authors gratefully acknowledge the support and scientific discussions with Jen Holt, Ivan Medvedev, Christopher Neese, Paul Helminger, and Doug Petkie. This work was partially supported by the Army's competitive in-house innovative laboratory program, the Defense Threat Reduction Agency (contract number HDTRA1-09-1-0031), and DARPA.

TABLE CAPTIONS

TABLE I. Molecular constants of methyl fluoride.

TABLE II. Molecular constants of methyl chloride.

TABLE III. Branching ratios for transitions involving ground state/stretching and $l$-doubled bending vibrational states. [30]

TABLE IV. Optically pumped far-infrared laser coincidences for methyl fluoride and methyl chloride.

TABLE V. Atmospheric pressure DR parameters for $^{12}CH_3F$ and $^{13}CH_3F$.

TABLE VI. Abbreviated table of atmospheric pressure DR parameters for $CH_3^{35}Cl$. The complete table may be found in the attached appendix.

FIGURE CAPTIONS

FIG. 1. Infrared spectra of a) $CO_2$ laser [23], b) $CH_3F$, and c) $CH_3Cl$ [28, 38] between 900 – 1100 cm$^{-1}$. Rotational THz spectra of d) J = 8 - 9 in $V_3 = 1$ of $CH_3F$ and e) J = 8 - 9 in $V_6 = 1$ of $CH_3Cl$.

FIG.2. (color online) The calculated $\Pi(v_j)$ for each $Q_K(12)$ transition of $^{12}CH_3F$ pumped by the 9P(20) $CO_2$ laser line, overlaid by the HWHM linewidth of IR ro-vibrational transitions of $CH_3F$ (continuous) and $CH_3Cl$ (dashed) from 1 – 760 Torr.

FIG. 3. (color online) Energy level diagram depicting the IR transition (black dashed arrow) and THz rotational transitions (red and blue solid arrows) connecting the pumped states $J_P$, $K_P$ and $J'_P, K'_P$ (left). The two pairs of THz DR signatures induced by the IR pump are illustrated for the $V_i$ (dotted red) and $V_0$ (dashed blue) vibrational levels, the sum of which for the three types of IR transition (P, Q, R-type) produces the measurable DR signature (right).

FIG. 4. a) (color online) Aggregate DR signature for the $^{12}CH_3F$ transitions pumped by the 9P(20) laser line, plotted above the constituent labeled components from Table V (to scale). Energy level diagram depicting both infrared pump transitions and rotational transitions connected to the pumped states. b) Aggregate DR signature for the $CH_3^{35}Cl$ transitions pumped by the 9P(26) laser line, plotted above the constituent components (to scale) labeled center.

Energy level diagram depicting both infrared pump transitions and rotational transitions connected to the pumped states. The transitions $^PR_0(24)$, $^RR_0(24)$ and $^RP_7(31)$ from Table VI are not depicted because they occur at higher frequencies.

FIG. 5. Calculated DR signatures for all laser concidences with a) $^{12}CH_3F$ and b) $^{13}CH_3F$. The coincident $CO_2$ laser lines are listed on the right.

FIG. 6. Calculated DR signatures for all laser concidences with $CH_3^{35}Cl$. The coincident $CO_2$ laser lines are listed on the right.

FIG. 7. Calculated DR signatures for all laser concidences with $CH_3^{37}Cl$. The coincident $CO_2$ laser lines are listed on the right.

TABLE I[*]

| | [12]CH$_3$F [34] | | [13]CH$_3$F [35, 36] | |
|---|---|---|---|---|
| | V$_3$=0 | V$_3$=1 | V$_3$=0 | V$_3$=1 |
| A (MHz) | 155352.70 (36) | 155058.474 (19) | 155365.345 (40) | 155076.4480 (309) |
| B (MHz) | 25536.14980 (54) | 25197.51060 (87) | 24862.665 (2) | 24542.131 (3) |
| D$_j$ (kHz) | 60.2214 (20) | 56.8868 (33) | 57.72435 (1130) | 55.06939 (1050) |
| D$_{jk}$ (kHz) | 439.8156 (240) | 518.2386 (471) | 424.83 (11) | 477.773 (140) |
| D$_k$ (kHz) | 2108.0 (75) | 2014.9000 (3358) | 2137.293 (1881) | 2071.5 (39) |
| H$_j$ (Hz) | -0.03214 (93) | -0.19320 (29) | -0.02130 (728) | -0.1200 (66) |
| H$_{jjk}$ (Hz) | 1.9470 (275) | 16.07690 (5156) | 1.5360 (872) | 9.7025 (1510) |
| H$_{jkk}$ (Hz) | 22.3610 (1601) | -96.44947 (48566) | 21.500 (764) | -38.421 (1130) |
| H$_k$ (Hz) | 0 | 0.1197871 (22754) | 150.4 (2471) | 231.99 (138) |
| L$_j$ (mHz) | 0 | 0.01020 (87) | 0 | 0 |
| L$_{jjjk}$ (mHz) | 0 | -1.7468 (187) | 0 | -0.7732 (594) |
| L$_{jjkk}$ (mHz) | 0 | 29.5648 (2138) | 0 | 13.300 (504) |
| L$_{jkkk}$ (mHz) | 0 | -173.2365 (14660) | 0 | -89.64 (270) |
| L$_k$ (mHz) | 0 | 0 | 0 | 0 |
| Vibrational Energy (cm$^{-1}$) | - | 1048.610701 (10) | - | 1027.49320 (2) |
| Dipole (debye) [31] | 1.857 | 1.857 | 1.857 | 1.857 |
| Dipole Derivative (debye) [31] | - | -0.271 | - | -0.271 |

---

[*] Uncertainties in the least significant digit(s) are in parentheses.

TABLE II[†]

|  | $^{12}CH_3^{35}Cl$ [37] | | $^{12}CH_3^{37}Cl$ [37] | |
| --- | --- | --- | --- | --- |
|  | $V_6=0$ | $V_6=1$ | $V_6=0$ | $V_6=1$ |
| A (MHz) | 156051.1 | 156810.71 (2) | 156440 | 157197.02 (3) |
| B (MHz) | 13292.8752 (63) | 13243.8840 (69) | 13088.1660 (93) | 13039.967 (10) |
| $D_j$ (kHz) | 18.1010 (39) | 18.1360 (45) | 17.5630 (78) | 17.5960 (87) |
| $D_{jk}$ (kHz) | 198.77 (13) | 203.37 (14) | 193.48 (13) | 197.85 (28) |
| $D_k$ (kHz) | 2653.1 | 2735.2 (1) | 2501.0 | 2583.0 (1) |
| $H_j$ (Hz) | -0.00845 (840) | -0.01010 (96) | -0.0121 (20) | -0.0140 (23) |
| $H_{ijk}$ (Hz) | 0.312 (33) | 0.375 (36) | 0.312 (33) | 0 |
| $H_{ikk}$ (Hz) | 9.53 (66) | 5.16 (72) | 8.39 (66) | 4.08 (190) |
| $H_k$ (Hz) | 0 | 0 | 0 | 0 |
| $A_\zeta$ (MHz) | 0 | 39275.246 (30) | 0 | 39560.346 (33) |
| q (MHz) | 0 | 3.639735 (250) | 0 | 3.58087 (33) |
| $q_i$ (kHz) | 0 | -0.01724 (10) | 0 | -0.017850 (17) |
| $q_{ii}$ (Hz) | 0 | 0 | 0 | 0 |
| $\eta_i$ (MHz) | 0 | 0.467059 (78) | 0 | 0.457777 (60) |
| $\eta_k$ (MHz) | 0 | 4.02759 (100) | 0 | 3.4039 (15) |
| $\eta_{ik}$ (kHz) | 0 | 0.12860 (57) | 0 | 0.1267 (10) |
| $\eta_{ii}$ (Hz) | 0 | -0.6601 (250) | 0 | 0 |
| $\eta_{kk}$ (kHz) | 0 | -0.1020 (45) | 0 | -0.1250 (75) |
| Vibrational Energy (cm$^{-1}$) | - | 1018.07110 (7) | - | 1017.69545 (1) |
| Dipole (Debye) [31] | 1.8989 | 1.8989 | 1.8989 | 1.8989 |
| Dipole Derivative (Debye) [31] | - | -0.0388 | - | -0.0388 |

[†] Uncertainties in the least significant digit(s) are in parentheses.

TABLE III

| Transition | Branching Ratio | l-Doubled Transition | Branching Ratio |
|---|---|---|---|
| $P_k(J)$ $\Delta J = -1,$ $\Delta K = 0$ | $\dfrac{J^2 - K^2}{J(2J+1)}$ | $^R P_k(J)$ | $\dfrac{(K-J)(K-J+1)}{4J(2J+1)}$ |
| | | $^P P_k(J)$ | $\dfrac{(-K-J)(-K-J+1)}{4J(2J+1)}$ |
| $Q_k(J)$ $\Delta J = 0,$ $\Delta K = 0$ | $\dfrac{K^2}{J(J+1)}$ | $^R Q_k(J)$ | $\dfrac{(J-K)(J+K+1)}{4J(J+1)}$ |
| | | $^P Q_k(J)$ | $\dfrac{(J+K)(J-K+1)}{4J(J+1)}$ |
| $R_k(J)$ $\Delta J = 1,$ $\Delta K = 0$ | $\dfrac{(J+1)^2 - K^2}{(J+1)(2J+1)}$ | $^R R_k(J)$ | $\dfrac{(J+K+1)(J+K+2)}{4(J+1)(2J+1)}$ |
| | | $^P R_k(J)$ | $\dfrac{(J-K+1)(J-K+2)}{4(J+1)(2J+1)}$ |

TABLE IV

| Molecule | Laser Line | IR Transition | IR Transition Offset (MHz) | $R_0^-$ (GHz) | $R_0^+$ (GHz) | $R_i^-$ (GHz) | $R_i^+$ (GHz) |
|---|---|---|---|---|---|---|---|
| $^{12}CH_3F$ | 9P(20) | $Q_2(12)$ | 39.78 [25] | 612.409 | 663.365 | 604.297 | 654.582 |
| $^{13}CH_3F$ | 9P(32) | $R_3(4)$ | -24.25 [25] | 198.855 [36] | 248.559 [36] | 245.350 [36] | 294.406 [36] |
| $^{12}CH_3^{35}Cl$ | 9P(26) | $^RQ_3(6)$ | 20 [26] | 159.477 | 186.050 | 158.894 | 185.370 |
| $^{12}CH_3^{35}Cl$ | 9R(12) | $^RR_6(11)$ | -30 [26] | 292.190 | 318.733 | 317.569 | 344.009 |
| $^{12}CH_3^{37}Cl$ | 9P(38) | $^RP_3(13)$ | -50 [26] | 340.093 | 366.227 | 312.806 | 338.850 |

TABLE V[‡]

| Laser Line | IR Transition | IR Transition Offset (GHz) | $R_0^-$ $\nu_j$ (GHz) $\Delta\alpha(\nu_j)/\varepsilon$ (km$^{-1}$kJ$^{-1}$) $\Pi(\nu_j)$ | $R_0^+$ $\nu_j$ (GHz) $\Delta\alpha(\nu_j)/\varepsilon$ (km$^{-1}$kJ$^{-1}$) $\Pi(\nu_j)$ | $R_i^-$ $\nu_j$ (GHz) $\Delta\alpha(\nu_j)/\varepsilon$ (km$^{-1}$kJ$^{-1}$) $\Pi(\nu_j)$ | $R_i^+$ $\nu_j$ (GHz) $\Delta\alpha(\nu_j)/\varepsilon$ (km$^{-1}$kJ$^{-1}$) $\Pi(\nu_j)$ |
|---|---|---|---|---|---|---|
| $^{12}CH_3F$ | | | | | | |
| 9P(18) | $Q_1(1)$ | -2.135 {1} | - | 102.141{1} | - | 100.786{1} |
| | | | - | -4.43 | - | 4.38 |
| | | | - | 0.27 | - | 0.25 |
| 9P(20) | $Q_{1:7}(12)$ | 0.042 {2} | 612.072 {6} | 663.000{6} | 603.900{6} | 654.152{6} |
| | | | 61.03 | -69.14 | -60.22 | 68.22 |
| | | | 3.06 | 3.04 | 3.05 | 3.04 |
| | $Q_{13}(13)$ | -1.198 {13} | - | 712.291 {13} | - | 702.391 {13} |
| | | | - | -1.61 | - | 1.59 |
| | | | - | 1.03 | - | 0.80 |
| | $P_0(1)$ | 1.585 {0} | 51.072 {0} | 102.143 {0} | - | 50.395 {0} |
| | | | 7.64 | -10.18 | - | 7.54 |
| | | | 1.90 | 1.00 | - | 1.93 |
| 9P(22) | $Q_{14:16}(18)$ | 0.0007 {15} | 914.385 {15} | 965.016 {15} | 901.451 {15} | 951.385 {15} |
| | | | 1.85 | -2.41 | -1.83 | 2.37 |
| | | | 1.77 | 1.74 | 1.74 | 1.72 |
| $^{13}CH_3F$ | | | | | | |
| 9P(08) | $R_{0:16}(20)$ | 0.535 {10} | 992.507 {3} | 1041.930 {3} | 1028.550 {3} | 1077.320 {3} |
| | | | 300.50 | -315.75 | -311.70 | 326.72 |
| | | | 4.49 | 4.62 | 4.60 | 4.46 |
| 9P(32) | $R_{0:4}(4)$ | -0.025 {3} | 198.887 {0} | 248.598 {0} | 245.394 {0} | 294.458 {0} |
| | | | 112.30 | -136.47 | -134.71 | 158.70 |
| | | | 2.92 | 4.40 | 4.38 | 3.55 |

[‡] The K quantum number of the transition closest to the $CO_2$ laser line is listed in {brackets} beside the IR transition offset frequency. The K quantum number of the transition with the lowest $\Pi(\nu)$ is listed in {brackets} beside the THz DR frequencies, and the values of $\Delta\alpha/\varepsilon$ and $\Pi(\nu)$ are only listed for these transitions.

TABLE VI

| Laser Line | IR Transition | IR Transition Offset (GHz) | $R_0^-$ $\nu_j$ (GHz) $\Delta\alpha(\nu_j)/\varepsilon$ (km$^{-1}$kJ$^{-1}$) $\Pi(\nu_j)$ | $R_0^+$ $\nu_j$ (GHz) $\Delta\alpha(\nu_j)/\varepsilon$ (km$^{-1}$kJ$^{-1}$) $\Pi(\nu_j)$ | $R_i^-$ $\nu_j$ (GHz) $\Delta\alpha(\nu_j)/\varepsilon$ (km$^{-1}$kJ$^{-1}$) $\Pi(\nu_j)$ | $R_i^+$ $\nu_j$ (GHz) $\Delta\alpha(\nu_j)/\varepsilon$ (km$^{-1}$kJ$^{-1}$) $\Pi(\nu_j)$ |
|---|---|---|---|---|---|---|
| 09P(26) | $^RQ_3(4)$ | 1.09 | 106.324 | 132.902 | - | 132.416 |
| | | | 0.072 | -0.128 | - | 0.072 |
| | | | 1.008 | -0.500 | - | 0.763 |
| | $^RQ_3(5)$ | 0.60 | 132.902 | 159.477 | 132.416 | 158.894 |
| | | | 0.263 | -0.364 | -0.148 | 0.268 |
| | | | 0.243 | -0.150 | -0.371 | 0.136 |
| | $^RQ_3(6)$ | 0.02 | 159.477 | 186.050 | 158.894 | 185.370 |
| | | | 0.533 | -0.669 | -0.394 | 0.550 |
| | | | 0.102 | -0.676 | -0.092 | 0.724 |
| | $^RR_2(7)$ | -1.28 | 186.064 | 212.636 | 211.858 | 238.329 |
| | | | 0.457 | -0.528 | -0.482 | 0.557 |
| | | | 0.990 | 1.709 | 1.825 | 0.951 |
| | $^RQ_3(7)$ | -0.66 | 186.050 | 212.620 | 185.370 | 211.843 |
| | | | 0.760 | -0.905 | -0.624 | 0.787 |
| | | | 0.595 | 0.998 | -0.637 | -1.115 |
| | $^RQ_3(8)$ | -1.44 | 212.620 | 239.187 | 211.843 | 238.312 |
| | | | 0.815 | -0.942 | -0.709 | 0.847 |
| | | | -1.108 | -0.244 | 1.238 | 0.630 |
| | $^RP_4(8)$ | 0.69 | 212.598 | 239.162 | 185.351 | 211.821 |
| | | | 0.041 | -0.049 | -0.024 | 0.033 |
| | | | -22.055 | -4.862 | -16.765 | -26.324 |
| | $^RR_0(24)$ | 1.04 | 637.057 | 663.512 | 661.071 | 687.418 |
| | | | 2.316 | -2.410 | -2.397 | 2.491 |
| | | | 0.970 | 1.425 | 1.430 | 0.971 |
| | $^PR_0(24)$ | 1.13 | 637.057 | 663.512 | 661.075 | 687.422 |
| | | | 2.244 | -2.335 | -2.323 | 2.414 |
| | | | 1.001 | 1.471 | 1.476 | 1.002 |
| | $^RP_7(31)$ | -0.31 | 821.399 | 847.750 | 792.123 | 818.390 |
| | | | 0.162 | -0.168 | -0.153 | 0.159 |
| | | | 1.345 | 0.988 | 0.986 | 1.359 |

Figure 1

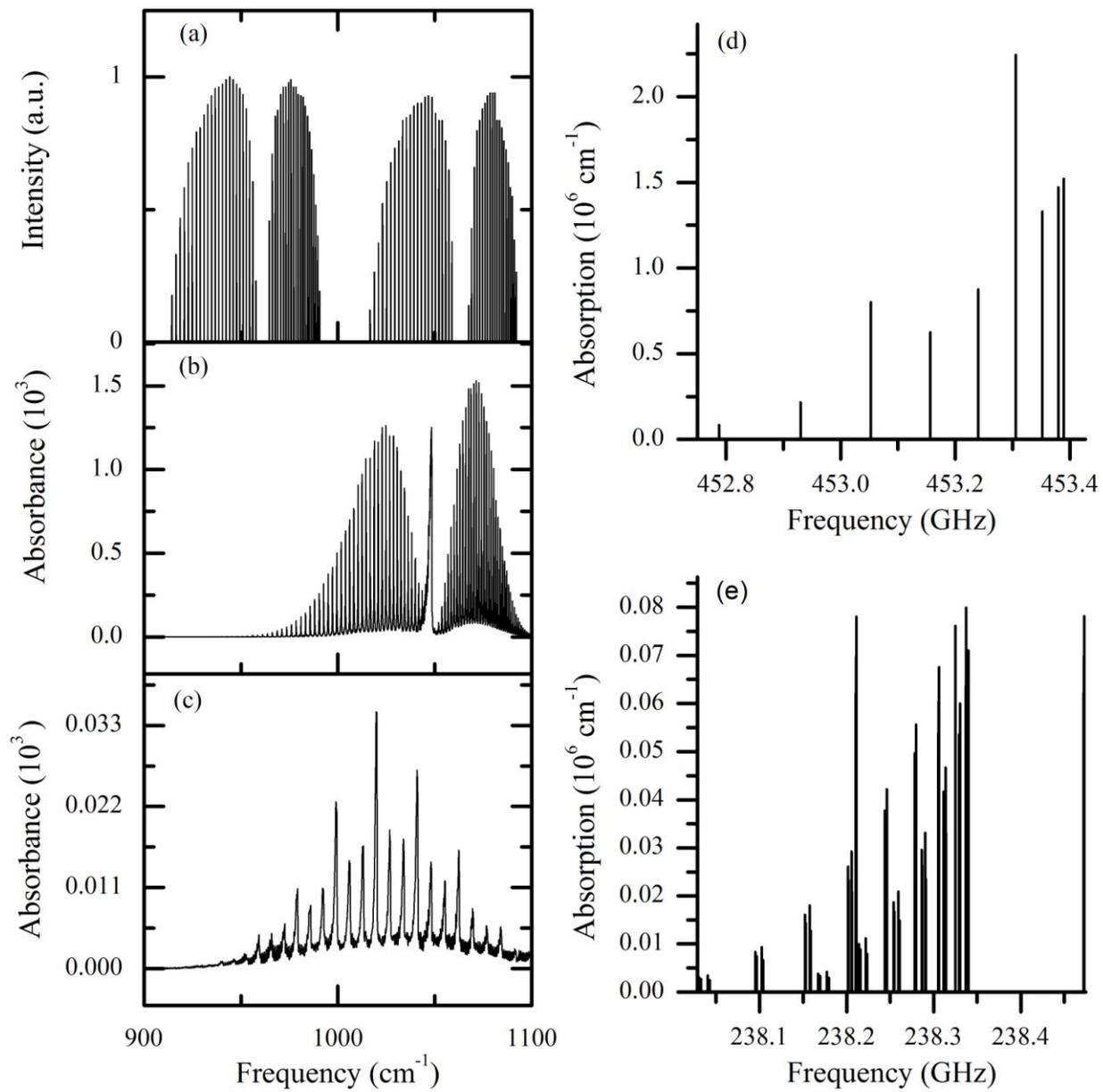

Figure 2

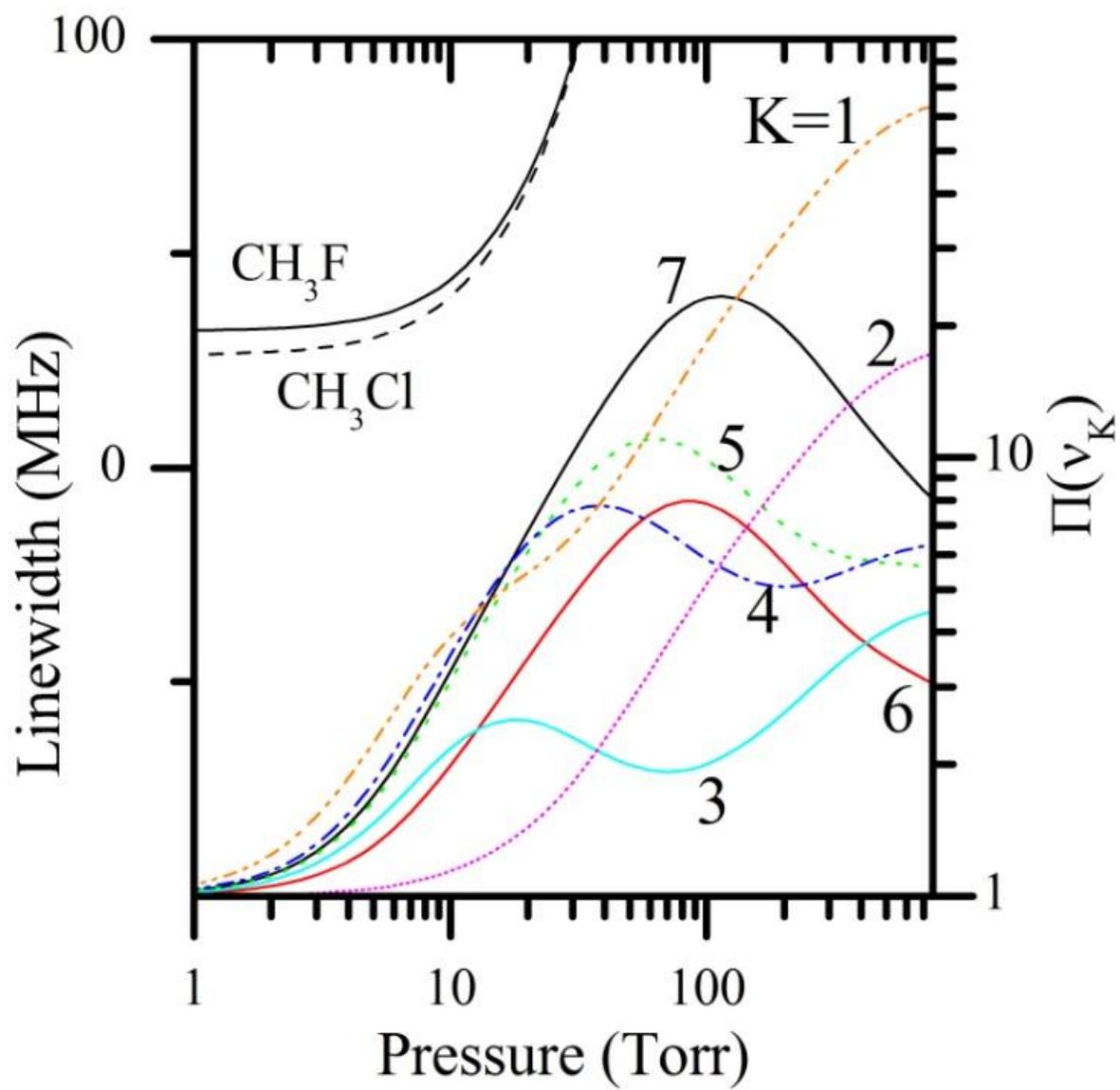

Figure 3

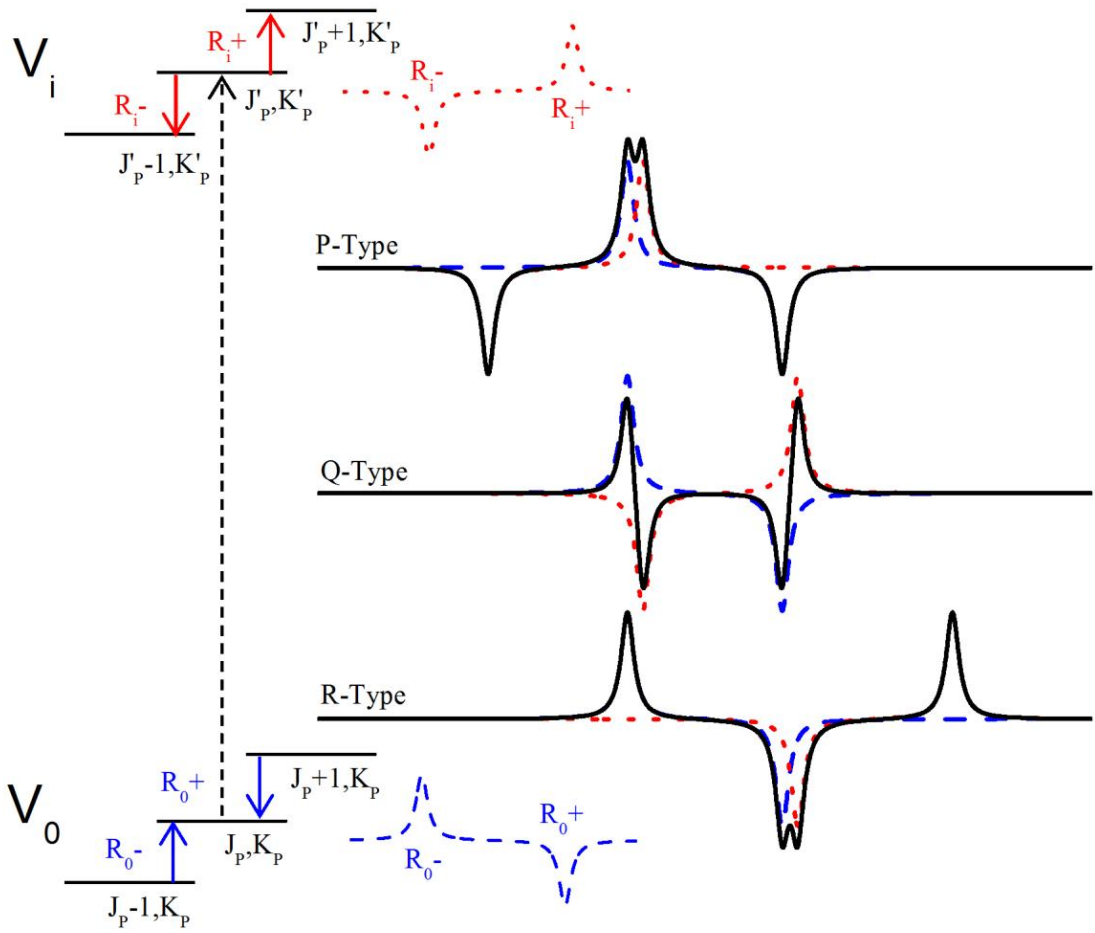

Figure 4

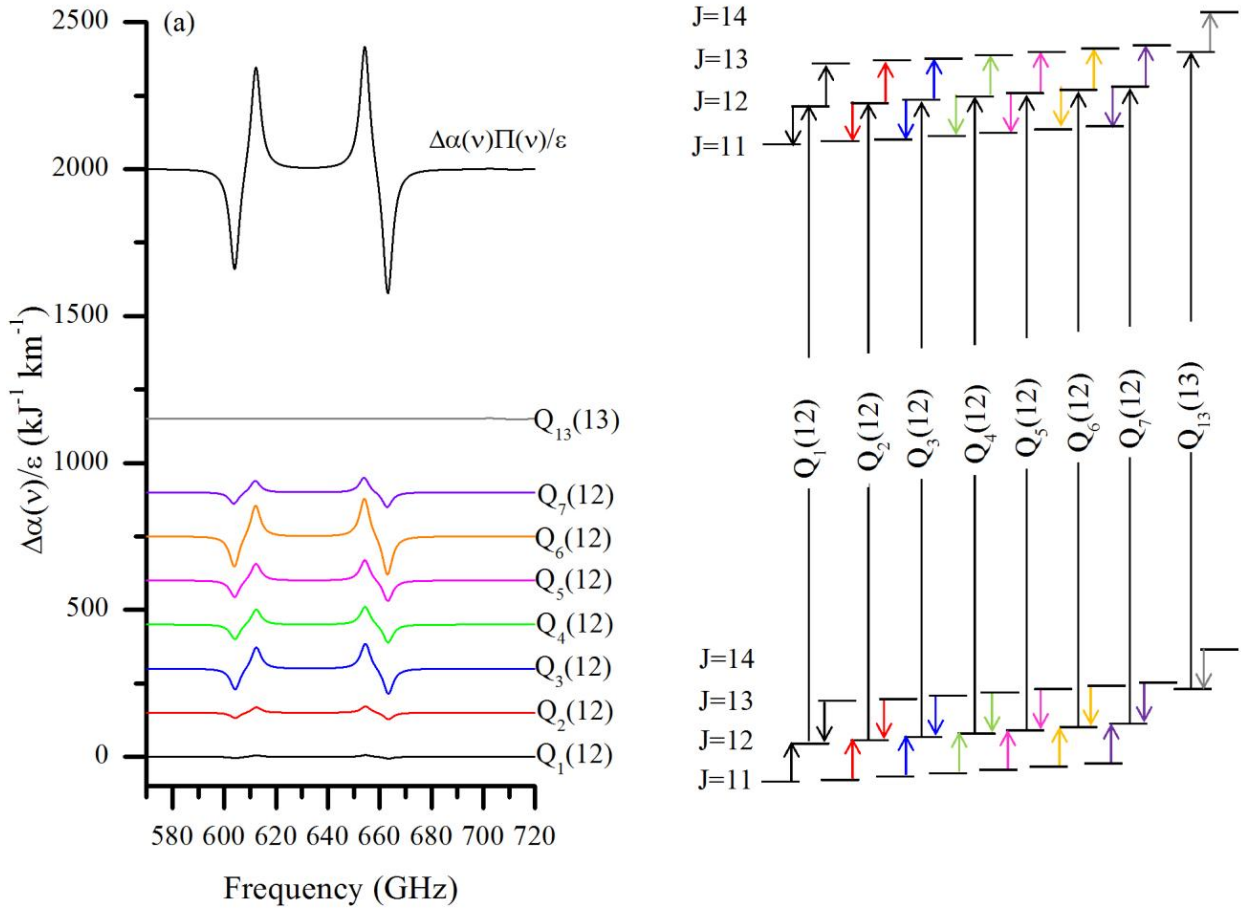

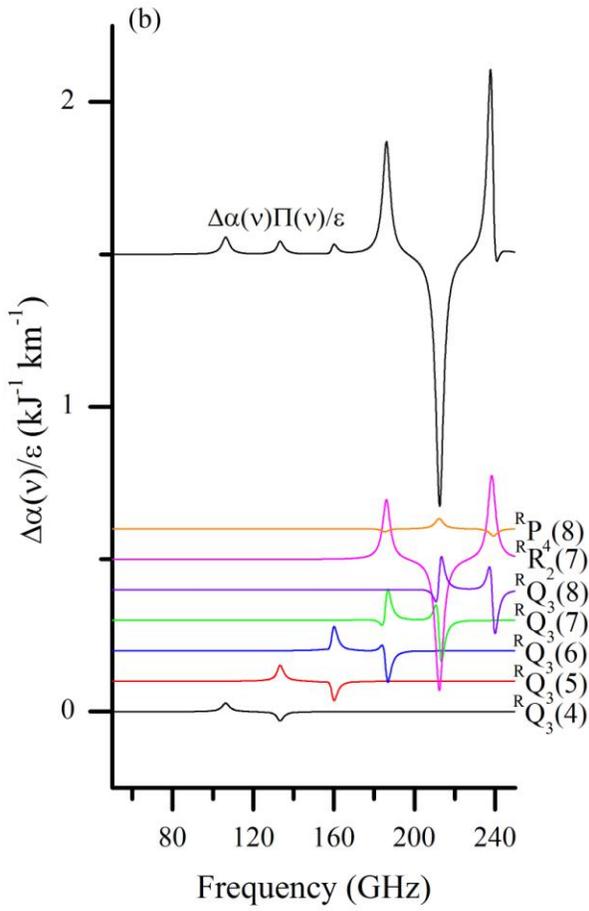 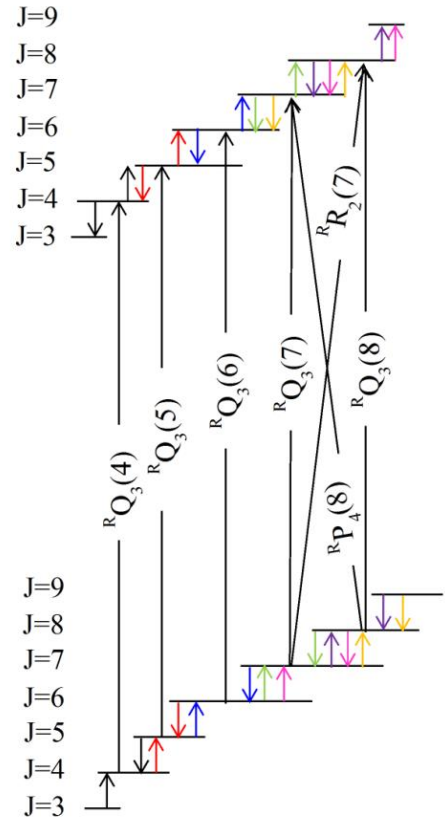

Figure 5

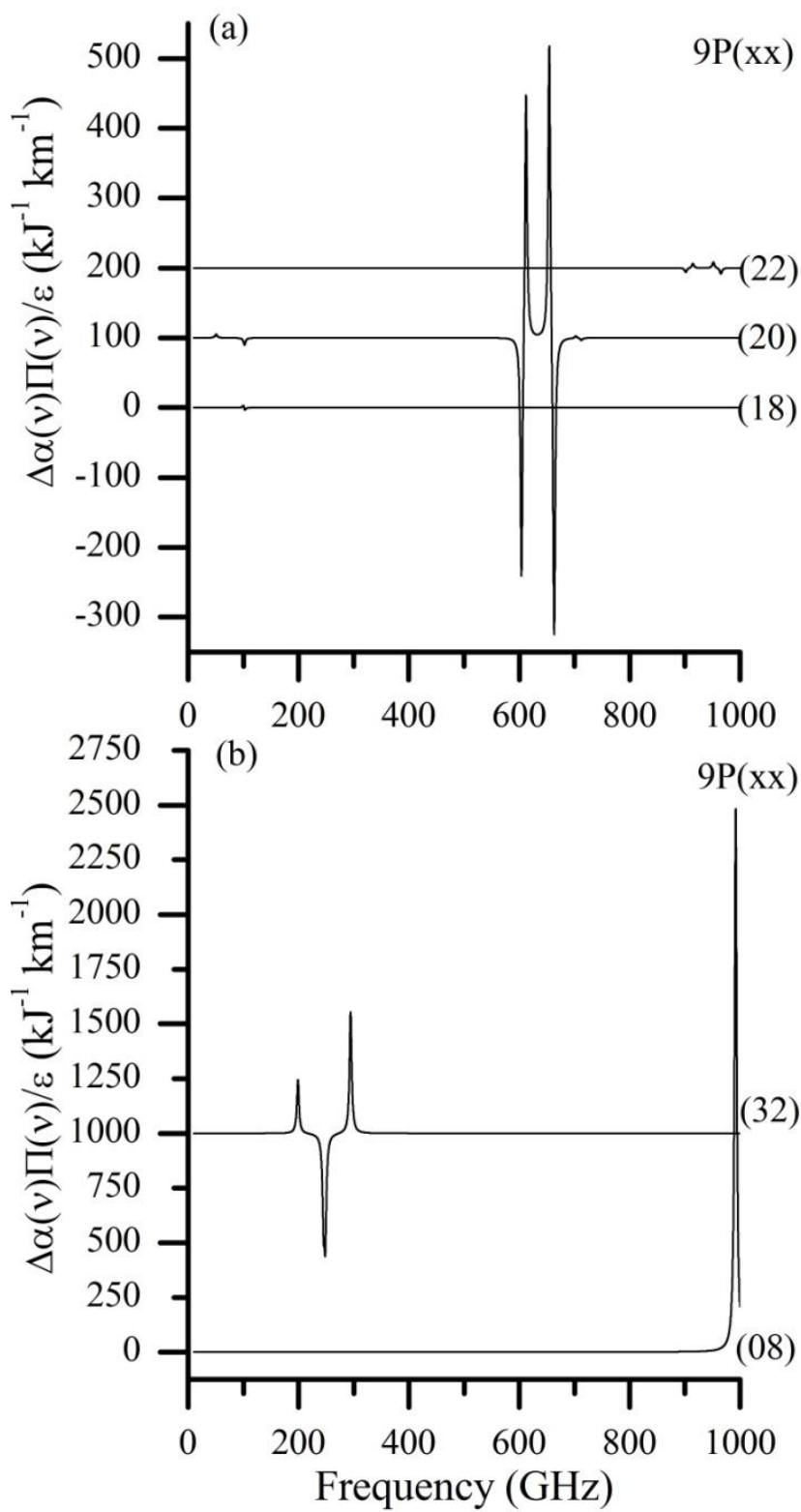

Figure 6

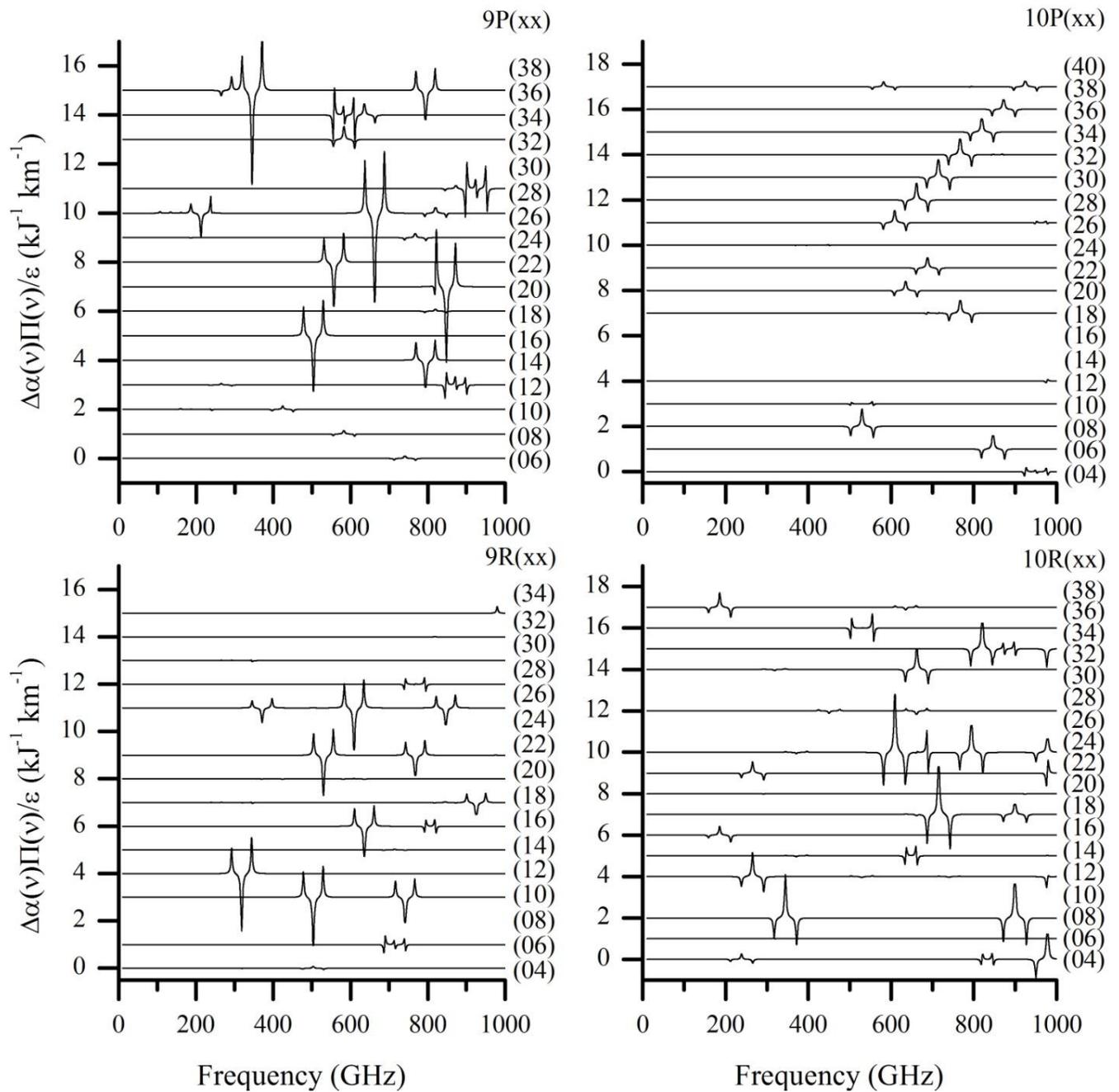

Figure 7

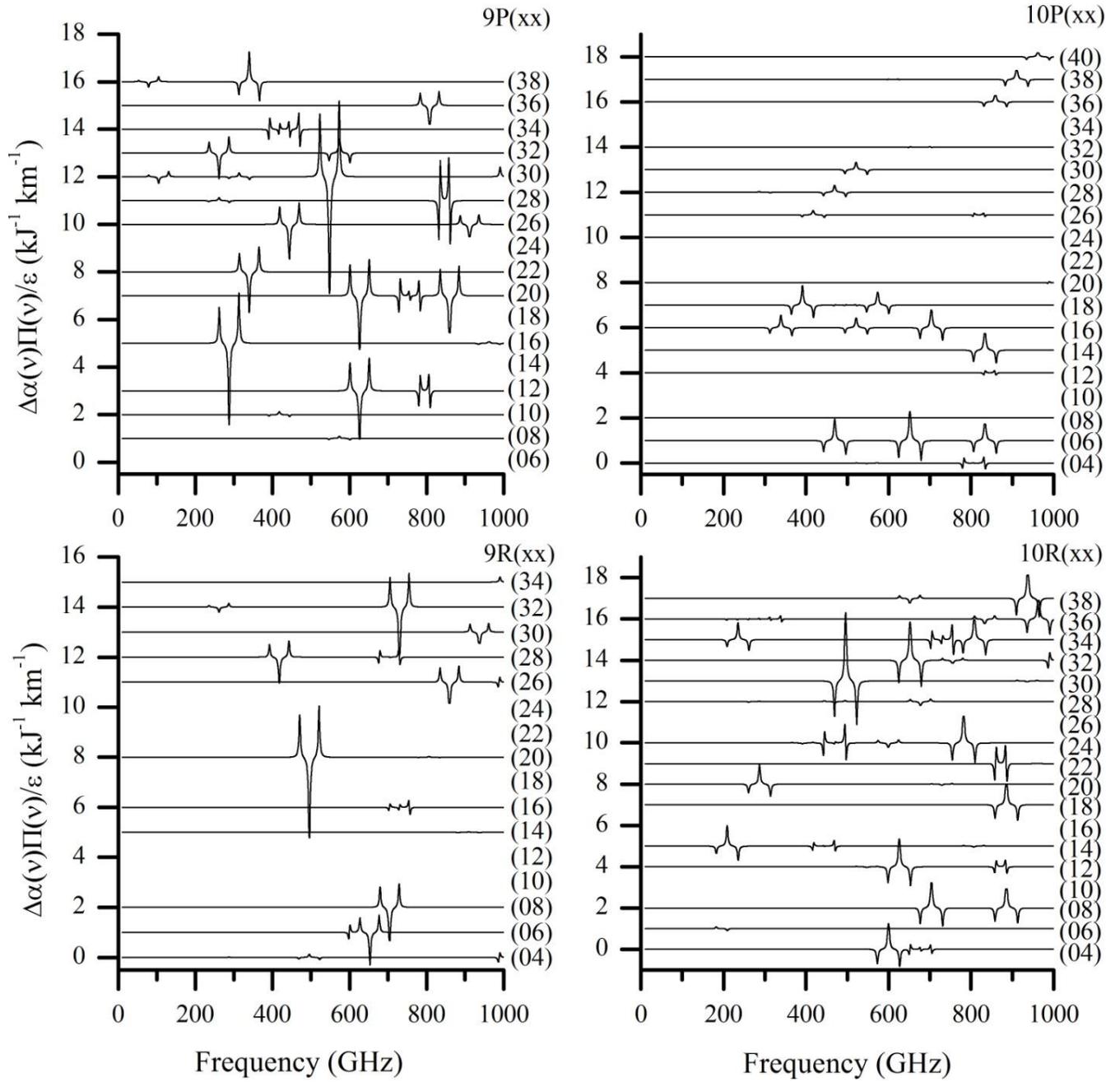

## Appendix

CH$_3^{35}$Cl

| Laser Line | IR Transition | IR Transition Offset (GHz) | R$_0$-<br>$\nu_j$ (GHz)<br>$\Delta\alpha(\nu_j)/\varepsilon$ (km$^{-1}$kJ$^{-1}$)<br>$\Pi(\nu_j)$ | R$_0$+<br>$\nu_j$ (GHz)<br>$\Delta\alpha(\nu_j)/\varepsilon$ (km$^{-1}$kJ$^{-1}$)<br>$\Pi(\nu_j)$ | R$_i$-<br>$\nu_j$ (GHz)<br>$\Delta\alpha(\nu_j)/\varepsilon$ (km$^{-1}$kJ$^{-1}$)<br>$\Pi(\nu_j)$ | R$_i$+<br>$\nu_j$ (GHz)<br>$\Delta\alpha(\nu_j)/\varepsilon$ (km$^{-1}$kJ$^{-1}$)<br>$\Pi(\nu_j)$ |
|---|---|---|---|---|---|---|
| 09P06 | $^RP_9(28)$ | -2.20 | 741.915<br>0.071<br>1.389 | 768.292<br>-0.074<br>0.989 | 712.907<br>-0.066<br>0.993 | 739.199<br>0.069<br>1.417 |
| 09P08 | $^RP_8(22)$ | -0.72 | 583.558<br>0.090<br>1.500 | 610.009<br>-0.096<br>0.989 | 555.062<br>-0.081<br>0.996 | 581.425<br>0.087<br>1.552 |
| 09P10 | $^RQ_5(6)$ | -0.75 | 159.439<br>0.027<br>1.002 | 186.006<br>-0.050<br>-0.229 | 0.000<br>0.000<br>0.000 | 185.327<br>0.027<br>0.321 |
|  | $^RQ_5(7)$ | -1.43 | 186.006<br>0.072<br>0.158 | 212.570<br>-0.102<br>0.003 | 185.327<br>-0.039<br>-0.222 | 211.793<br>0.073<br>-0.002 |
|  | $^RQ_5(8)$ | -2.20 | 212.570<br>0.101<br>-0.003 | 239.130<br>-0.127<br>0.301 | 211.793<br>-0.072<br>0.002 | 238.256<br>0.102<br>-0.089 |
|  | $^RP_7(16)$ | -0.29 | 424.765<br>0.085<br>1.629 | 451.272<br>-0.092<br>0.989 | 396.796<br>-0.070<br>1.009 | 423.213<br>0.078<br>1.743 |
| 09P12 | $^RP_6(10)$ | -0.96 | 265.642<br>0.038<br>1.713 | 292.190<br>-0.045<br>0.990 | 238.218<br>-0.021<br>1.122 | 264.672<br>0.030<br>2.131 |
|  | $^RQ_5(32)$ | 1.03 | 848.053<br>0.775<br>0.638 | 874.399<br>-0.799<br>0.197 | 844.944<br>-0.763<br>0.714 | 871.192<br>0.788<br>0.395 |
|  | $^RQ_5(33)$ | -2.17 | 874.399<br>0.453<br>-0.348 | 900.731<br>-0.467<br>0.807 | 871.192<br>-0.447<br>-0.697 | 897.427<br>0.461<br>0.611 |
|  | $^RP_9(33)$ | 2.15 | 873.671<br>0.063<br>-1.157 | 899.981<br>-0.065<br>4.336 | 844.242<br>-0.059<br>9.307 | 870.469<br>0.061<br>5.004 |
| 09P14 | $^RR_1(29)$ | 1.68 | 769.208 | 795.604 | 792.680 | 818.965 |
|       |                |       | 0.759    | -0.785   | -0.780   | 0.805    |
|       |                |       | 0.985    | 1.359    | 1.364    | 0.986    |
|       | $^RR_0(38)$    | -1.64 | 1006.280 | 1032.540 | 1028.730 | 1054.880 |
|       |                |       | 0.342    | -0.351   | -0.349   | 0.358    |
|       |                |       | 1.014    | 1.282    | 1.284    | 1.014    |
|       | $^PR_0(38)$    | -1.50 | 1006.280 | 1032.540 | 1028.740 | 1054.880 |
|       |                |       | 0.363    | -0.372   | -0.370   | 0.380    |
|       |                |       | 0.956    | 1.209    | 1.212    | 0.957    |
| 09P16 | $^RR_2(18)$    | 1.03  | 478.093  | 504.602  | 502.754  | 529.158  |
|       |                |       | 1.265    | -1.335   | -1.312   | 1.382    |
|       |                |       | 0.986    | 1.577    | 1.598    | 0.987    |
| 09P18 | $^RP_8(31)$    | -2.08 | 821.216  | 847.561  | 791.947  | 818.208  |
|       |                |       | 0.058    | -0.060   | -0.054   | 0.056    |
|       |                |       | 1.344    | 0.989    | 0.991    | 1.362    |
| 09P20 | $^RR_0(31)$    | -1.07 | 822.000  | 848.370  | 845.244  | 871.503  |
|       |                |       | 0.814    | -0.839   | -0.836   | 0.861    |
|       |                |       | 1.477    | 1.836    | 0.924    | 1.003    |
|       | $^PR_0(31)$    | -0.95 | 822.000  | 848.370  | 845.248  | 871.506  |
|       |                |       | 0.845    | -0.872   | -0.868   | 0.895    |
|       |                |       | 1.422    | 1.767    | 0.891    | 0.966    |
|       | $^RQ_4(31)$    | 0.67  | 821.803  | 848.167  | 818.789  | 845.056  |
|       |                |       | 1.161    | -1.199   | -1.145   | 1.183    |
|       |                |       | 2.040    | 2.584    | 0.161    | -1.221   |
| 09P22 | $^RR_1(20)$    | 1.57  | 531.128  | 557.622  | 555.576  | 581.963  |
|       |                |       | 1.048    | -1.099   | -1.087   | 1.139    |
|       |                |       | 0.986    | 1.532    | 1.544    | 0.986    |
| 09P24 | $^RP_4(6)$     | -1.40 | 159.461  | 186.031  | 0.000    | 158.878  |
|       |                |       | 0.004    | -0.006   | 0.000    | 0.002    |
|       |                |       | 1.652    | 0.992    | 0.000    | 2.983    |
|       | $^RP_7(29)$    | 1.67  | 768.658  | 795.035  | 739.551  | 765.843  |
|       |                |       | 0.116    | -0.121   | -0.109   | 0.114    |
|       |                |       | 1.375    | 0.989    | 0.990    | 1.394    |
| 09P26 | $^RQ_3(4)$     | 1.09  | 106.324  | 132.902  | 0.000    | 132.416  |
|       |                |       | 0.072    | -0.128   | 0.000    | 0.072    |
|       |                |       | 1.008    | -0.500   | 0.000    | 0.763    |
|       | $^RQ_3(5)$     | 0.60  | 132.902  | 159.477  | 132.416  | 158.894  |
|       |                |       | 0.263    | -0.364   | -0.148   | 0.268    |
|       |                |       | 0.243    | -0.150   | -0.371   | 0.136    |
|       | $^RQ_3(6)$     | 0.02  | 159.477  | 186.050  | 158.894  | 185.370  |
|  |  |  | 0.533 | -0.669 | -0.394 | 0.550 |
|  |  |  | 0.102 | -0.672 | -0.092 | 0.719 |
|  | $^RR_2(7)$ | -1.28 | 186.064 | 212.636 | 211.858 | 238.329 |
|  |  |  | 0.457 | -0.528 | -0.482 | 0.557 |
|  |  |  | 0.984 | 1.706 | 1.821 | 0.950 |
|  | $^RQ_3(7)$ | -0.66 | 186.050 | 212.620 | 185.370 | 211.843 |
|  |  |  | 0.760 | -0.905 | -0.624 | 0.787 |
|  |  |  | 0.592 | 0.996 | -0.633 | -1.113 |
|  | $^RQ_3(8)$ | -1.44 | 212.620 | 239.187 | 211.843 | 238.312 |
|  |  |  | 0.815 | -0.942 | -0.709 | 0.847 |
|  |  |  | -1.106 | -0.244 | 1.235 | 0.630 |
|  | $^RP_4(8)$ | 0.69 | 212.598 | 239.162 | 185.351 | 211.821 |
|  |  |  | 0.041 | -0.049 | -0.024 | 0.033 |
|  |  |  | -22.010 | -4.866 | -16.650 | -26.263 |
|  | $^RR_0(24)$ | 1.04 | 637.057 | 663.512 | 661.071 | 687.418 |
|  |  |  | 1.158 | -1.205 | -1.198 | 1.245 |
|  |  |  | 0.970 | 1.425 | 1.430 | 0.971 |
|  | $^PR_0(24)$ | 1.13 | 637.057 | 663.512 | 661.075 | 687.422 |
|  |  |  | 1.122 | -1.167 | -1.161 | 1.207 |
|  |  |  | 1.001 | 1.471 | 1.476 | 1.002 |
|  | $^RP_7(31)$ | -0.31 | 821.399 | 847.750 | 792.123 | 818.390 |
|  |  |  | 0.162 | -0.168 | -0.153 | 0.159 |
|  |  |  | 1.345 | 0.988 | 0.989 | 1.361 |
| 09P28 | $^RP_7(33)$ | -1.86 | 874.087 | 900.409 | 844.644 | 870.883 |
|  |  |  | 0.089 | -0.092 | -0.085 | 0.088 |
|  |  |  | 1.281 | -9.134 | 0.993 | 1.305 |
|  | $^RQ_3(34)$ | 1.45 | 900.946 | 927.270 | 897.637 | 923.863 |
|  |  |  | 1.698 | -1.748 | -1.682 | 1.731 |
|  |  |  | 0.631 | 0.193 | 0.696 | 0.197 |
|  | $^RQ_3(35)$ | -1.96 | 927.270 | 953.578 | 923.863 | 950.073 |
|  |  |  | 1.255 | -1.291 | -1.243 | 1.279 |
|  |  |  | -0.268 | 0.706 | -0.275 | 0.699 |
| 09P32 | $^RP_5(22)$ | -0.90 | 583.897 | 610.363 | 555.383 | 581.761 |
|  |  |  | 0.342 | -0.359 | -0.315 | 0.332 |
|  |  |  | 1.506 | 0.989 | 0.991 | 1.537 |
| 09P34 | $^RQ_2(21)$ | 0.69 | 557.597 | 584.080 | 555.553 | 581.939 |
|  |  |  | 2.201 | -2.305 | -2.168 | 2.273 |
|  |  |  | 0.454 | 0.122 | 0.437 | 0.121 |
|  | $^RQ_2(22)$ | -1.45 | 584.080 | 610.554 | 581.939 | 608.316 |
|       |              |       | 1.708    | -1.785   | -1.684   | 1.761    |
|       |              |       | -0.165   | 0.570    | -0.163   | 0.325    |
|       | $^RP_5(24)$  | 1.33  | 636.819  | 663.264  | 608.130  | 634.488  |
|       |              |       | 0.305    | -0.319   | -0.284   | 0.298    |
|       |              |       | 1.458    | 0.990    | -2.218   | 1.480    |
| 09P36 | $^RP_3(11)$  | -1.75 | 292.308  | 318.861  | 264.778  | 291.239  |
|       |              |       | 0.330    | -0.364   | -0.273   | 0.309    |
|       |              |       | 1.785    | -4.271   | 0.996    | 1.897    |
|       | $^RR_0(12)$  | -0.33 | 318.904  | 345.456  | 344.187  | 370.635  |
|       |              |       | 0.968    | -1.045   | -1.035   | 1.112    |
|       |              |       | 0.803    | 1.747    | 1.761    | 0.990    |
|       | $^PR_0(12)$  | -0.28 | 318.904  | 345.456  | 344.190  | 370.639  |
|       |              |       | 0.973    | -1.051   | -1.041   | 1.118    |
|       |              |       | 0.798    | 1.737    | 1.752    | 0.984    |
|       | $^PR_2(29)$  | -0.70 | 769.174  | 795.569  | 792.633  | 818.916  |
|       |              |       | 0.816    | -0.844   | -0.843   | 0.871    |
|       |              |       | 0.985    | 1.359    | 1.360    | 0.986    |
|       | $^PR_3(38)$  | 0.72  | 1006.150 | 1032.410 | 1028.580 | 1054.720 |
|       |              |       | 0.652    | -0.669   | -0.669   | 0.686    |
|       |              |       | 0.984    | 1.244    | 1.244    | 0.985    |
| 09R04 | $^RP_8(11)$  | 0.87  | 292.068  | 318.600  | 264.562  | 291.002  |
|       |              |       | 0.004    | -0.005   | -0.001   | 0.002    |
|       |              |       | 1.652    | 0.991    | 1.337    | 2.315    |
|       | $^RP_9(19)$  | -1.77 | 504.024  | 530.495  | 475.800  | 502.182  |
|       |              |       | 0.049    | -0.053   | -0.042   | 0.046    |
|       |              |       | 1.559    | 0.989    | 1.008    | 1.655    |
| 09R06 | $^RR_6(6)$   | -0.10 | 0.000    | 185.975  | 185.297  | 211.759  |
|       |              |       | 0.000    | -0.328   | 0.000    | 0.327    |
|       |              |       | 0.000    | 0.992    | 0.000    | 0.992    |
|       | $^RQ_7(26)$  | 0.67  | 689.452  | 715.866  | 686.930  | 713.246  |
|       |              |       | 0.635    | -0.663   | -0.618   | 0.646    |
|       |              |       | 0.563    | 0.238    | 0.537    | 0.234    |
|       | $^RQ_7(27)$  | -1.95 | 715.866  | 742.268  | 713.246  | 739.551  |
|       |              |       | 0.388    | -0.404   | -0.378   | 0.394    |
|       |              |       | -0.407   | 0.597    | -0.399   | 0.576    |
| 09R10 | $^RR_5(18)$  | -0.13 | 477.943  | 504.444  | 502.600  | 528.996  |
|       |              |       | 1.144    | -1.216   | -1.172   | 1.245    |
|       |              |       | 0.986    | 1.567    | 1.611    | 0.987    |
|       | $^RR_4(27)$  | 1.69  | 716.218  | 742.633  | 739.912  | 766.217  |
| | | | 0.714 | -0.741 | -0.730 | 0.757 |
| | | | 0.985 | 1.389 | 1.402 | 0.986 |
| 09R12 | $^RR_6(11)$ | -0.03 | 292.190 | 318.733 | 317.569 | 344.009 |
| | | | 1.204 | -1.397 | -1.224 | 1.422 |
| | | | 0.986 | 1.681 | 1.889 | 0.987 |
| | $^RR_3(38)$ | -0.73 | 1006.150 | 1032.410 | 1028.610 | 1054.750 |
| | | | 0.883 | -0.906 | -0.899 | 0.922 |
| | | | 0.984 | 1.245 | 1.249 | 0.985 |
| 09R14 | $^RP_{11}(27)$ | 1.13 | 715.101 | 741.475 | 686.195 | 712.484 |
| | | | 0.016 | -0.017 | -0.015 | 0.016 |
| | | | 1.404 | 0.989 | 0.997 | 1.444 |
| 09R16 | $^RR_5(23)$ | -1.47 | 610.363 | 636.819 | 634.488 | 660.836 |
| | | | 0.781 | -0.817 | -0.798 | 0.834 |
| | | | 0.985 | 1.461 | 1.484 | 0.986 |
| | $^RQ_8(30)$ | -0.10 | 794.858 | 821.216 | 791.947 | 818.208 |
| | | | 0.399 | -0.414 | -0.390 | 0.405 |
| | | | 0.631 | 0.642 | 0.609 | 0.629 |
| 09R18 | $^RQ_8(9)$ | 1.21 | 238.991 | 265.532 | 0.000 | 264.562 |
| | | | 0.009 | -0.017 | 0.000 | 0.009 |
| | | | 1.008 | -0.736 | 0.000 | 0.814 |
| | $^RQ_8(10)$ | 0.24 | 265.532 | 292.068 | 264.562 | 291.002 |
| | | | 0.039 | -0.056 | -0.020 | 0.039 |
| | | | 0.317 | -0.099 | -0.351 | 0.042 |
| | $^RQ_8(11)$ | -0.83 | 292.068 | 318.600 | 291.002 | 317.437 |
| | | | 0.069 | -0.088 | -0.048 | 0.069 |
| | | | 0.080 | 0.081 | -0.034 | -0.002 |
| | $^RQ_8(12)$ | -1.99 | 318.600 | 345.126 | 317.437 | 343.866 |
| | | | 0.069 | -0.083 | -0.054 | 0.070 |
| | | | -0.103 | 0.361 | 0.003 | 0.082 |
| | $^RP_{12}(32)$ | -0.18 | 846.556 | 872.856 | 817.237 | 843.453 |
| | | | 0.021 | -0.022 | -0.019 | 0.020 |
| | | | 1.338 | 0.926 | 0.991 | 1.363 |
| | $^RR_4(34)$ | 2.00 | 900.852 | 927.173 | 923.768 | 949.976 |
| | | | 0.372 | -0.383 | -0.379 | 0.390 |
| | | | 0.984 | 1.289 | 1.296 | 0.985 |
| 09R20 | $^RP_{10}(15)$ | 1.09 | 397.949 | 424.443 | 370.092 | 396.495 |
| | | | 0.005 | -0.005 | -0.003 | 0.004 |
| | | | 1.617 | 0.990 | 1.085 | 1.911 |
| | $^RP_{11}(23)$ | -2.00 | 609.493 | 635.911 | 580.933 | 607.264 |
| | | | 0.009 | -0.010 | -0.008 | 0.008 |
| | | | 1.474 | 0.989 | 1.004 | 1.544 |
| 09R22 | $^RR_6(19)$ | 2.20 | 504.361 | 530.850 | 528.910 | 555.293 |
| | | | 0.982 | -1.044 | -1.003 | 1.065 |
| | | | 0.986 | 1.541 | 1.587 | 0.986 |
| | $^RR_5(28)$ | 1.68 | 742.533 | 768.933 | 766.116 | 792.405 |
| | | | 0.576 | -0.597 | -0.587 | 0.609 |
| | | | 0.985 | 1.372 | 1.386 | 0.986 |
| | $^RP_{13}(37)$ | 0.68 | 977.547 | 1003.760 | 947.824 | 973.957 |
| | | | 0.004 | -0.004 | -0.004 | 0.004 |
| | | | 1.270 | 0.987 | 0.989 | 1.288 |
| 09R26 | $^RR_7(13)$ | -1.93 | 345.203 | 371.730 | 370.372 | 396.796 |
| | | | 0.322 | -0.365 | -0.326 | 0.371 |
| | | | 0.986 | 1.646 | 1.809 | 0.987 |
| | $^RP_{11}(19)$ | 0.62 | 503.725 | 530.180 | 475.517 | 501.884 |
| | | | 0.008 | -0.009 | -0.006 | 0.007 |
| | | | 1.576 | 0.914 | 1.002 | 1.717 |
| | $^RR_6(22)$ | 1.94 | 583.801 | 610.263 | 608.031 | 634.385 |
| | | | 1.076 | -1.132 | -1.098 | 1.154 |
| | | | 0.985 | 1.479 | 1.510 | 0.986 |
| | $^RR_5(31)$ | -1.51 | 821.693 | 848.053 | 844.944 | 871.192 |
| | | | 0.502 | -0.519 | -0.511 | 0.528 |
| | | | 0.985 | 1.328 | 1.338 | 0.985 |
| 09R28 | $^RQ_9(28)$ | 1.61 | 741.915 | 768.292 | 739.199 | 765.480 |
| | | | 0.382 | -0.398 | -0.370 | 0.387 |
| | | | 0.599 | -0.053 | 0.573 | -0.044 |
| | $^RQ_9(29)$ | -1.20 | 768.292 | 794.657 | 765.480 | 791.747 |
| | | | 0.430 | -0.448 | -0.418 | 0.436 |
| | | | 0.049 | 0.629 | 0.041 | 0.610 |
| 09R30 | $^RQ_9(10)$ | 0.97 | 265.464 | 291.994 | 0.000 | 290.928 |
| | | | 0.013 | -0.024 | 0.000 | 0.013 |
| | | | 1.006 | -0.661 | 0.000 | 0.662 |
| | $^RQ_9(11)$ | -0.10 | 291.994 | 318.519 | 290.928 | 317.356 |
| | | | 0.052 | -0.075 | -0.027 | 0.052 |
| | | | 0.307 | -0.024 | -0.307 | 0.008 |
| | $^RQ_9(12)$ | -1.26 | 318.519 | 345.039 | 317.356 | 343.779 |
| | | | 0.079 | -0.101 | -0.055 | 0.079 |
| | | | 0.023 | 0.401 | -0.008 | 0.020 |
| 09R32 | $^RP_{13}(31)$ | 0.51 | 819.940 | 846.244 | 790.714 | 816.934 |
| | | | 0.005 | -0.005 | -0.005 | 0.005 |
| | | | 1.341 | 0.989 | 0.996 | 1.373 |
| 09R34 | $^RR_5(37)$ | -1.47 | 979.636 | 1005.910 | 1002.210 | 1028.370 |
| | | | 0.288 | -0.296 | -0.292 | 0.300 |
| | | | 0.984 | 1.255 | 1.261 | 0.985 |
| 10P04 | $^PQ_9(35)$ | 1.54 | 926.276 | 952.556 | 922.788 | 948.967 |
| | | | 0.260 | -0.268 | -0.263 | 0.271 |
| | | | 0.698 | 0.201 | 0.703 | 0.209 |
| | $^PQ_9(36)$ | -2.05 | 952.556 | 978.820 | 948.967 | 975.132 |
| | | | 0.191 | -0.197 | -0.193 | 0.199 |
| | | | -0.282 | 0.721 | -0.294 | 0.725 |
| 10P06 | $^PP_5(32)$ | 1.46 | 848.053 | 874.399 | 818.641 | 844.902 |
| | | | 0.437 | -0.451 | -0.426 | 0.440 |
| | | | 1.332 | 0.989 | 0.988 | 1.328 |
| 10P08 | $^PP_7(20)$ | 1.44 | 530.747 | 557.222 | 502.385 | 528.770 |
| | | | 0.466 | -0.495 | -0.458 | 0.485 |
| | | | 1.578 | 0.988 | 0.983 | 1.531 |
| 10P10 | $^PQ_{10}(19)$ | 1.07 | 503.882 | 530.345 | 501.983 | 528.347 |
| | | | 0.121 | -0.132 | -0.129 | 0.140 |
| | | | 0.369 | -0.019 | 0.447 | -0.017 |
| | $^PQ_{10}(20)$ | -0.93 | 530.345 | 556.800 | 528.347 | 554.701 |
| | | | 0.137 | -0.149 | -0.146 | 0.156 |
| | | | 0.018 | 0.431 | 0.016 | 0.485 |
| 10P12 | $^PR_{14}(26)$ | -0.94 | 687.954 | 714.310 | 711.556 | 737.798 |
| | | | 0.002 | -0.002 | -0.002 | 0.003 |
| | | | 0.984 | 1.408 | 1.372 | 0.986 |
| | $^PR_{15}(34)$ | 1.05 | 898.070 | 924.309 | 920.714 | 946.835 |
| | | | 0.002 | -0.002 | -0.002 | 0.002 |
| | | | 0.982 | 1.285 | 1.269 | 0.933 |
| | $^PQ_{10}(37)$ | 0.39 | 978.544 | 1004.780 | 974.842 | 1000.980 |
| | | | 0.095 | -0.098 | -0.096 | 0.099 |
| | | | 0.725 | 0.731 | 0.726 | 0.741 |
| 10P18 | $^PQ_{11}(26)$ | 1.92 | 688.716 | 715.101 | 686.105 | 712.390 |
| | | | 0.063 | -0.067 | -0.066 | 0.069 |
| | | | 0.548 | -0.217 | 0.586 | -0.276 |
| | $^PQ_{11}(27)$ | -0.79 | 715.101 | 741.475 | 712.390 | 738.663 |
| | | | 0.095 | -0.100 | -0.098 | 0.103 |
| | | | 0.152 | 2.721 | 0.193 | -2.627 |
| | $^PP_7(29)$ | -1.07 | 768.658 | 795.035 | 739.497 | 765.788 |
| | | | 0.393 | -0.408 | -0.384 | 0.399 |
| | | | 1.378 | 0.989 | 0.899 | 1.366 |
| | $^PR_{16}(35)$ | -0.95 | 923.890 | 950.102 | 946.378 | 972.470 |
| | | | 0.000 | 0.000 | 0.000 | 0.000 |
| | | | 0.941 | 1.284 | 1.272 | 0.972 |
| 10P20 | $^PP_8(24)$ | 1.98 | 636.449 | 662.879 | 607.727 | 634.068 |
| | | | 0.260 | -0.273 | -0.256 | 0.268 |
| | | | 1.478 | 0.988 | 0.984 | 1.447 |
| 10P22 | $^PP_8(26)$ | -1.45 | 689.298 | 715.706 | 660.398 | 686.718 |
| | | | 0.300 | -0.314 | -0.294 | 0.308 |
| | | | 1.435 | 0.988 | 0.985 | 1.412 |
| 10P24 | $^PQ_{12}(14)$ | 1.58 | 371.206 | 397.690 | 369.795 | 396.177 |
| | | | 0.011 | -0.015 | -0.015 | 0.020 |
| | | | -0.041 | -0.090 | 0.499 | -0.363 |
| | $^PQ_{12}(15)$ | 0.06 | 397.690 | 424.167 | 396.177 | 422.553 |
| | | | 0.028 | -0.036 | -0.035 | 0.043 |
| | | | 0.050 | 0.043 | 0.202 | 0.100 |
| | $^PQ_{12}(16)$ | -1.55 | 424.167 | 450.636 | 422.553 | 448.922 |
| | | | 0.028 | -0.034 | -0.033 | 0.039 |
| | | | -0.055 | 0.263 | -0.128 | 0.446 |
| | $^PR_{16}(28)$ | 2.00 | 740.003 | 766.313 | 763.314 | 789.507 |
| | | | 0.000 | 0.000 | 0.000 | 0.000 |
| | | | 0.984 | 1.366 | 1.333 | 0.986 |
| 10P26 | $^PR_{15}(18)$ | -0.26 | 476.530 | 502.953 | 501.003 | 527.315 |
| | | | 0.000 | 0.000 | 0.000 | 0.000 |
| | | | 0.886 | 1.870 | 1.609 | 0.563 |
| | $^PP_9(23)$ | 2.21 | 609.854 | 636.288 | 581.220 | 607.564 |
| | | | 0.338 | -0.358 | -0.335 | 0.353 |
| | | | 1.503 | 0.988 | 0.982 | 1.458 |
| | $^PQ_{12}(36)$ | 0.72 | 951.668 | 977.907 | 948.033 | 974.171 |
| | | | 0.068 | -0.070 | -0.069 | 0.071 |
| | | | 0.716 | 0.723 | 0.720 | 0.735 |
| 10P28 | $^PP_9(25)$ | 1.17 | 662.712 | 689.124 | 633.898 | 660.222 |
| | | | 0.490 | -0.514 | -0.483 | 0.507 |
| | | | 1.457 | 0.988 | 0.983 | 1.423 |
| 10P30 | $^PP_9(27)$ | 0.52 | 715.525 | 741.915 | 686.535 | 712.836 |
| | | | 0.538 | -0.562 | -0.530 | 0.554 |
| | | | 1.415 | 0.989 | 0.984 | 1.390 |
| | $^PR_{18}(36)$ | -2.03 | 949.160 | 975.331 | 971.441 | 997.490 |
| | | | 0.000 | 0.000 | 0.000 | 0.000 |
| | | | 0.927 | 1.281 | 1.268 | 0.960 |
| 10P32 | $^PP_9(29)$ | 0.25 | 768.292 | 794.657 | 739.126 | 765.403 |
| | | | 0.497 | -0.517 | -0.489 | 0.509 |
| | | | 1.378 | 0.989 | 0.985 | 1.359 |
| | $^PQ_{13}(32)$ | 0.22 | 846.244 | 872.534 | 842.997 | 869.186 |
| | | | 0.026 | -0.027 | -0.027 | 0.028 |
| | | | 0.649 | 0.679 | 0.694 | 0.688 |
| 10P34 | $^PR_{16}(16)$ | 0.94 | 0.000 | 449.895 | 448.138 | 474.453 |
| | | | 0.000 | 0.000 | 0.000 | 0.000 |
| | | | 0.000 | 1.139 | 0.000 | 0.919 |
| | $^PP_9(31)$ | 0.39 | 821.008 | 847.347 | 791.668 | 817.920 |
| | | | 0.424 | -0.440 | -0.417 | 0.432 |
| | | | 1.345 | 0.989 | 0.986 | 1.331 |
| | $^PR_{18}(31)$ | 2.07 | 818.081 | 844.325 | 840.963 | 867.088 |
| | | | 0.000 | 0.000 | 0.000 | 0.000 |
| | | | 6355.580 | 1569.010 | 402.501 | -38.064 |
| 10P36 | $^PP_9(33)$ | 0.93 | 873.671 | 899.981 | 844.158 | 870.383 |
| | | | 0.317 | -0.328 | -0.311 | 0.322 |
| | | | 1.316 | 0.989 | 0.986 | 1.305 |
| | $^PR_{19}(36)$ | -0.79 | 948.650 | 974.807 | 970.882 | 996.916 |
| | | | 0.000 | 0.000 | 0.000 | 0.000 |
| | | | -9.867 | 4.155 | 4.550 | -0.221 |
| 10P38 | $^PP_{11}(22)$ | -0.34 | 583.065 | 609.493 | 554.519 | 580.856 |
| | | | 0.134 | -0.144 | -0.135 | 0.144 |
| | | | 1.535 | 0.988 | 0.976 | 1.456 |
| | $^PQ_{14}(30)$ | 0.89 | 793.307 | 819.614 | 790.246 | 816.451 |
| | | | 0.013 | -0.013 | -0.013 | 0.014 |
| | | | 0.631 | 0.303 | 0.655 | 0.313 |
| | $^PQ_{14}(31)$ | -2.27 | 819.614 | 845.908 | 816.451 | 842.642 |
| | | | 0.007 | -0.007 | -0.007 | 0.008 |
| | | | -0.569 | 0.682 | -0.587 | 0.665 |
| | $^PP_9(35)$ | 1.88 | 926.276 | 952.556 | 896.593 | 922.788 |
| | | | 0.182 | -0.188 | -0.179 | 0.184 |
| | | | 1.290 | 0.989 | 0.987 | 1.281 |
| 10R04 | $^PP_7(9)$ | 1.87 | 239.044 | 265.591 | 211.703 | 238.154 |
| | | | 0.125 | -0.179 | -0.156 | 0.201 |
| | | | 2.286 | 0.989 | 0.884 | 1.466 |
| | $^PR_{10}(12)$ | -0.49 | 318.429 | 344.942 | 343.643 | 370.049 |
| | | | 0.002 | -0.003 | -0.004 | 0.004 |
| | | | 0.905 | 1.958 | 1.594 | 0.985 |
| | $^PQ_8(31)$ | -0.23 | 821.216 | 847.561 | 818.138 | 844.383 |
| | | | 0.381 | -0.395 | -0.386 | 0.400 |
| | | | 0.643 | 0.655 | 0.649 | 0.666 |
| | $^PP_3(37)$ | -0.50 | 979.870 | 1006.150 | 950.046 | 976.240 |
| | | | 0.966 | -0.991 | -0.940 | 0.965 |
| | | | 1.270 | 0.989 | 0.988 | 1.270 |
| 10R06 | $^PQ_8(9)$ | 0.11 | 238.991 | 265.532 | 238.098 | 264.540 |
| | | | 0.021 | -0.041 | -0.040 | 0.057 |
| | | | -0.627 | 0.038 | 0.538 | -0.081 |
| | $^PQ_8(10)$ | -0.89 | 265.532 | 292.068 | 264.540 | 290.977 |
| | | | 0.049 | -0.070 | -0.069 | 0.088 |
| | | | -0.032 | -0.006 | 0.068 | 0.070 |
| | $^PQ_8(11)$ | -1.98 | 292.068 | 318.600 | 290.977 | 317.410 |
| | | | 0.056 | -0.072 | -0.071 | 0.085 |
| | | | 0.008 | 0.070 | -0.087 | 0.336 |
| 10R08 | $^PP_6(13)$ | 0.76 | 345.270 | 371.802 | 317.550 | 343.988 |
| | | | 1.178 | -1.312 | -1.173 | 1.298 |
| | | | 1.814 | 0.988 | 0.974 | 1.666 |
| | $^PP_3(34)$ | 0.64 | 900.946 | 927.270 | 871.372 | 897.612 |
| | | | 1.267 | -1.304 | -1.231 | 1.268 |
| | | | 1.305 | 0.989 | 0.988 | 1.305 |
| 10R12 | $^PP_6(10)$ | -1.50 | 265.642 | 292.190 | 238.203 | 264.657 |
| | | | 0.617 | -0.741 | -0.648 | 0.758 |
| | | | 1.994 | 0.989 | 0.951 | 1.648 |
| | $^PR_{10}(19)$ | -0.21 | 503.882 | 530.345 | 528.347 | 554.701 |
| | | | 0.027 | -0.030 | -0.031 | 0.034 |
| | | | 0.984 | 1.583 | 1.519 | 0.987 |
| | $^PR_{11}(27)$ | 1.37 | 715.101 | 741.475 | 738.663 | 764.924 |
| | | | 0.018 | -0.019 | -0.020 | 0.021 |
| | | | 0.984 | 1.391 | 1.370 | 0.986 |
| | $^PQ_7(37)$ | 1.69 | 979.286 | 1005.550 | 975.621 | 1001.780 |
| | | | 0.217 | -0.223 | -0.218 | 0.224 |
| | | | 0.044 | -2.216 | 2.427 | 2.278 |
| | $^PP_2(38)$ | 0.85 | 1006.220 | 1032.480 | 976.319 | 1002.500 |
| | | | 0.424 | -0.435 | -0.412 | 0.423 |
| | | | 1.164 | 1.268 | 1.277 | 1.338 |
| | $^PQ_7(38)$ | -2.08 | 1005.550 | 1031.790 | 1001.780 | 1027.930 |
| | | | 0.165 | -0.170 | -0.166 | 0.171 |
| | | | 2.990 | 3.052 | -3.074 | 0.275 |
| 10R14 | $^PR_9(13)$ | 1.78 | 345.039 | 371.553 | 370.160 | 396.568 |
| | | | 0.016 | -0.019 | -0.022 | 0.025 |
| | | | 0.983 | 1.816 | 1.622 | 0.988 |
| | $^PQ_7(24)$ | 0.12 | 636.591 | 663.027 | 634.217 | 660.554 |
| | | | 0.719 | -0.754 | -0.734 | 0.768 |
| | | | 0.512 | 0.531 | 0.529 | 0.550 |
| | $^PR_{12}(37)$ | 0.71 | 977.907 | 1004.130 | 1000.290 | 1026.400 |
| | | | 0.017 | -0.017 | -0.018 | 0.018 |
| | | | 0.984 | 1.246 | 1.240 | 0.985 |
| 10R16 | $^PP_6(7)$ | -1.64 | 185.975 | 212.535 | 0.000 | 185.285 |
| | | | 0.180 | -0.335 | 0.000 | 0.407 |
| | | | 2.867 | 0.988 | 0.000 | 1.304 |
| 10R18 | $^PR_{10}(24)$ | -1.68 | 636.109 | 662.525 | 660.026 | 686.330 |
| | | | 0.028 | -0.029 | -0.030 | 0.032 |
| | | | 0.943 | 1.735 | 1.638 | -39.627 |
| | $^PP_3(27)$ | -1.42 | 716.293 | 742.711 | 687.315 | 713.647 |
| | | | 1.616 | -1.676 | -1.561 | 1.621 |
| | | | 1.414 | 0.989 | 0.971 | 1.412 |
| | $^PP_2(34)$ | -2.20 | 901.014 | 927.339 | 871.443 | 897.685 |
| | | | 0.376 | -0.387 | -0.364 | 0.375 |
| | | | 1.305 | 0.989 | 0.988 | 1.307 |
| 10R20 | $^PR_8(10)$ | -0.11 | 265.532 | 292.068 | 290.977 | 317.410 |
| | | | 0.006 | -0.008 | -0.011 | 0.013 |
| | | | 0.979 | 2.007 | 1.633 | 0.989 |
| | $^PR_{11}(34)$ | 1.73 | 899.447 | 925.727 | 922.210 | 948.374 |
| | | | 0.013 | -0.014 | -0.014 | 0.014 |
| | | | 0.984 | 1.283 | 1.274 | 0.985 |
| 10R22 | $^PP_5(10)$ | -2.23 | 265.686 | 292.238 | 238.245 | 264.703 |
| | | | 0.296 | -0.343 | -0.298 | 0.340 |
| | | | 1.939 | 0.989 | 0.967 | 1.708 |
| | $^PQ_6(37)$ | -0.64 | 979.475 | 1005.740 | 975.821 | 1001.990 |
| | | | 0.866 | -0.890 | -0.869 | 0.893 |
| | | | 0.721 | 0.728 | 0.719 | 0.734 |
| 10R24 | $^PR_8(13)$ | -0.60 | 345.126 | 371.647 | 370.258 | 396.674 |
| | | | 0.029 | -0.034 | -0.037 | 0.042 |
| | | | 0.983 | 1.793 | 1.643 | 0.988 |
| | $^PP_3(23)$ | -1.46 | 610.509 | 636.971 | 581.881 | 608.255 |
| | | | 1.828 | -1.908 | -1.759 | 1.839 |
| | | | 1.494 | 0.560 | 0.988 | 1.489 |
| | $^PQ_6(24)$ | 1.41 | 636.714 | 663.155 | 634.347 | 660.689 |
| | | | 1.488 | -1.556 | -1.512 | 1.580 |
| | | | -0.741 | -0.036 | 1.051 | -0.045 |
| | $^PQ_6(25)$ | -1.05 | 663.155 | 689.586 | 660.689 | 687.021 |
| | | | 1.671 | -1.744 | -1.696 | 1.767 |
| | | | 0.034 | 0.553 | 0.042 | 0.564 |
| | $^PR_{10}(29)$ | 0.80 | 768.076 | 794.433 | 791.433 | 817.677 |
| | | | 0.040 | -0.042 | -0.043 | 0.045 |
| | | | -13.605 | -31.494 | -21.654 | -3.698 |
| | $^PP_2(30)$ | 0.15 | 795.569 | 821.951 | 766.336 | 792.633 |
| | | | 0.987 | -1.019 | -0.954 | 0.986 |
| | | | 1.319 | 0.979 | 0.962 | 1.303 |
| | $^PP_1(37)$ | -0.66 | 979.987 | 1006.270 | 950.171 | 976.369 |
| | | | 0.501 | -0.515 | -0.486 | 0.500 |
| | | | 1.270 | 0.989 | 0.988 | 1.272 |
| 10R28 | $^PR_8(16)$ | 0.99 | 424.670 | 451.171 | 449.485 | 475.880 |
| | | | 0.056 | -0.062 | -0.066 | 0.071 |
| | | | 0.985 | 1.671 | 1.593 | 0.987 |
| | $^PR_9(24)$ | 1.74 | 636.288 | 662.712 | 660.222 | 686.535 |
| | | | 0.100 | -0.106 | -0.109 | 0.114 |
| | | | 0.985 | 1.452 | 1.428 | 0.986 |
| 10R32 | $^PR_7(11)$ | 1.32 | 292.133 | 318.671 | 317.484 | 343.918 |
| | | | 0.025 | -0.030 | -0.034 | 0.038 |
| | | | 0.983 | 1.875 | 1.681 | 0.988 |
| | $^PP_2(25)$ | 2.16 | 663.472 | 689.915 | 634.671 | 661.027 |
| | | | 0.672 | -0.699 | -0.646 | 0.672 |
| | | | 1.452 | 0.989 | 0.988 | 1.452 |
| 10R34 | $^PP_1(31)$ | -0.21 | 821.988 | 848.357 | 792.673 | 818.958 |
| | | | 0.917 | -0.946 | -0.885 | 0.914 |
| | | | 1.345 | 0.420 | 0.988 | 1.348 |
| | $^PQ_5(32)$ | 1.00 | 848.053 | 874.399 | 844.902 | 871.150 |
| | | | 0.793 | -0.818 | -0.797 | 0.822 |
| | | | -0.502 | 0.308 | 0.999 | 0.302 |
| | $^PQ_5(33)$ | -2.25 | 874.399 | 900.731 | 871.150 | 897.383 |
| | | | 0.445 | -0.459 | -0.447 | 0.460 |
| | | | -0.567 | 0.685 | -0.555 | 0.681 |
| | $^RP_0(38)$ | 0.02 | 1006.280 | 1032.540 | 976.386 | 1002.570 |
| | | | 0.480 | -0.492 | -0.465 | 0.478 |
| | | | 1.257 | 0.986 | 0.986 | 1.259 |
| | $^PP_0(38)$ | 0.16 | 1006.280 | 1032.540 | 976.390 | 1002.570 |
| | | | 0.477 | -0.490 | -0.463 | 0.475 |
| | | | 1.262 | 0.991 | 0.991 | 1.265 |
| 10R36 | $^PQ_5(19)$ | 1.14 | 504.444 | 530.937 | 502.575 | 528.970 |
| | | | 1.057 | -1.119 | -1.081 | 1.141 |
| | | | 0.388 | -0.032 | 0.415 | -0.029 |
| | $^PQ_5(20)$ | -0.83 | 530.937 | 557.422 | 528.970 | 555.356 |
| | | | 1.199 | -1.265 | -1.223 | 1.287 |
| | | | 0.030 | 0.441 | 0.027 | 0.461 |
| 10R38 | $^PP_4(7)$ | 0.34 | 186.031 | 212.598 | 158.872 | 185.344 |
| | | | 0.372 | -0.469 | -0.386 | 0.469 |
| | | | 2.115 | 0.989 | 0.952 | 1.695 |
| | $^PR_8(23)$ | 2.17 | 610.009 | 636.449 | 634.068 | 660.398 |
| | | | 0.069 | -0.073 | -0.074 | 0.078 |
| | | | 0.985 | 1.474 | 1.450 | 0.986 |

CH$_3$$^{37}$Cl

| Laser Line | IR Transition | IR Transition Offset (GHz) | | R$_0$- $\nu_j$ (GHz) $\Delta\alpha(\nu_j)/\varepsilon$ (km$^{-1}$kJ$^{-1}$) $\Pi(\nu_j)$ | R$_0$+ $\nu_j$ (GHz) $\Delta\alpha(\nu_j)/\varepsilon$ (km$^{-1}$kJ$^{-1}$) $\Pi(\nu_j)$ | R$_i$- $\nu_j$ (GHz) $\Delta\alpha(\nu_j)/\varepsilon$ (km$^{-1}$kJ$^{-1}$) $\Pi(\nu_j)$ | R$_i$+ $\nu_j$ (GHz) $\Delta\alpha(\nu_j)/\varepsilon$ (km$^{-1}$kJ$^{-1}$) $\Pi(\nu_j)$ |
|---|---|---|---|---|---|---|---|
| 09P08 | 1.96 | | $^R$P$_8$(22) | 574.589 | 600.634 | 546.531 | 572.490 |
| | | | | 0.056 | -0.059 | -0.050 | 0.054 |
| | | | | 1.508 | 0.988 | 0.996 | 1.560 |
| 09P10 | -1.65 | | $^R$P$_6$(8) | 209.263 | 235.411 | - | 208.500 |
| | | | | 0.004 | -0.005 | - | 0.002 |
| | | | | 1.629 | 0.992 | - | 2.999 |
| | -0.76 | | $^R$P$_7$(16) | 418.231 | 444.331 | 390.694 | 416.705 |
| | | | | 0.076 | -0.082 | -0.063 | 0.070 |
| | | | | 1.636 | 0.989 | 1.008 | 1.750 |
| 09P12 | -1.01 | | $^R$R$_2$(23) | 601.166 | 627.224 | 624.925 | 650.878 |
| | | | | 1.236 | -1.289 | -1.273 | 1.327 |
| | | | | 0.985 | 1.473 | 1.485 | 0.986 |
| | -0.43 | | $^R$Q$_5$(30) | 783.106 | 809.077 | 780.235 | 806.109 |
| | | | | 1.028 | -1.063 | -1.011 | 1.047 |
| | | | | 0.622 | 0.633 | 0.605 | 0.625 |
| 09P16 | -0.57 | | $^R$R$_3$(10) | 261.658 | 287.808 | 286.756 | 312.806 |
| | | | | 1.695 | -1.888 | -1.763 | 1.963 |
| | | | | 0.985 | 1.755 | 1.866 | 0.987 |
| | -2.14 | | $^R$P$_9$(37) | 963.820 | 989.670 | 934.506 | 960.274 |
| | | | | 0.048 | -0.050 | -0.046 | 0.047 |
| | | | | 1.274 | 0.988 | 0.990 | 1.286 |
| 09P20 | -0.63 | | $^R$R$_1$(23) | 601.192 | 627.252 | 624.951 | 650.904 |
| | | | | 1.381 | -1.440 | -1.427 | 1.486 |
| | | | | 0.985 | 1.474 | 1.483 | 0.986 |
| | 1.18 | | $^R$Q$_4$(28) | 731.224 | 757.223 | 728.544 | 754.447 |
| | | | | 1.179 | -1.222 | -1.161 | 1.204 |
| | | | | 0.587 | 0.129 | 0.573 | 0.133 |
| | -1.60 | | $^R$Q$_4$(29) | 757.223 | 783.210 | 754.447 | 780.337 |
| | | | | 0.954 | -0.987 | -0.940 | 0.974 |

|  |  |  |  |  |  |  |
|---|---|---|---|---|---|---|
|  |  |  | -0.165 | 0.618 | -0.170 | 0.608 |
|  | -1.75 | $^RR_0(32)$ | 835.343 | 861.297 | 858.121 | 883.963 |
|  |  |  | 0.582 | -0.600 | -0.597 | 0.615 |
|  |  |  | 0.993 | 1.277 | 1.263 | 0.978 |
|  | -1.63 | $^PR_0(32)$ | 835.343 | 861.297 | 858.124 | 883.967 |
|  |  |  | 0.612 | -0.630 | -0.627 | 0.646 |
|  |  |  | 0.944 | 1.214 | 1.202 | 0.930 |
|  | 0.85 | $^RP_8(33)$ | 860.485 | 886.401 | 831.494 | 857.327 |
|  |  |  | 0.084 | -0.087 | -0.079 | 0.082 |
|  |  |  | -18.461 | -5.621 | -2.741 | -15.381 |
| 09P22 | -1.24 | $^RR_2(12)$ | 313.976 | 340.118 | 338.874 | 364.915 |
|  |  |  | 0.862 | -0.934 | -0.902 | 0.977 |
|  |  |  | 0.985 | 1.729 | 1.781 | 0.987 |
| 09P26 | -2.01 | $^RR_1(16)$ | 418.527 | 444.646 | 443.017 | 469.033 |
|  |  |  | 0.784 | -0.832 | -0.820 | 0.868 |
|  |  |  | 0.985 | 1.637 | 1.655 | 0.986 |
|  | 2.19 | $^PR_1(34)$ | 887.224 | 913.150 | 909.776 | 935.589 |
|  |  |  | 0.373 | -0.384 | -0.383 | 0.393 |
|  |  |  | 0.984 | 1.295 | 1.297 | 0.985 |
| 09P28 | -1.55 | $^RP_4(10)$ | 261.631 | 287.778 | 234.621 | 260.676 |
|  |  |  | 0.076 | -0.086 | -0.056 | 0.067 |
|  |  |  | 1.763 | 0.989 | 1.023 | 1.967 |
|  | -0.67 | $^RQ_3(32)$ | 835.232 | 861.183 | 832.164 | 858.018 |
|  |  |  | 2.592 | -2.672 | -2.564 | 2.645 |
|  |  |  | 0.650 | 0.660 | 0.639 | 0.657 |
| 09P30 | -1.45 | $^RR_2(3)$ | 78.523 | 104.695 | 104.312 | 130.387 |
|  |  |  | 0.101 | -0.172 | -0.100 | 0.178 |
|  |  |  | 0.983 | 1.553 | 2.652 | 0.989 |
|  | 2.17 | $^RP_4(12)$ | 313.920 | 340.058 | 286.728 | 312.774 |
|  |  |  | 0.103 | -0.114 | -0.085 | 0.096 |
|  |  |  | 1.736 | 0.988 | 1.004 | 1.861 |
|  | 0.14 | $^RR_0(20)$ | 522.965 | 549.053 | 547.036 | 573.018 |
|  |  |  | 1.422 | -1.491 | -1.483 | 1.552 |
|  |  |  | 0.982 | 1.537 | 1.544 | 0.983 |
| | 0.22 | $^PR_0(20)$ | 522.965 | 549.053 | 547.039 | 573.022 |
| | | | 1.415 | -1.483 | -1.475 | 1.543 |
| | | | 0.988 | 1.546 | 1.553 | 0.989 |
| | -0.38 | $^PR_2(38)$ | 990.793 | 1016.660 | 1012.880 | 1038.630 |
| | | | 0.421 | -0.432 | -0.431 | 0.442 |
| | | | 0.984 | 1.249 | 1.250 | 0.984 |
| 09P32 | -1.53 | $^RR_1(9)$ | 235.532 | 261.689 | 260.731 | 286.788 |
| | | | 0.527 | -0.583 | -0.564 | 0.621 |
| | | | 0.985 | 1.806 | 1.864 | 0.987 |
| | 0.13 | $^RP_5(22)$ | 574.919 | 600.979 | 546.844 | 572.818 |
| | | | 0.388 | -0.407 | -0.358 | 0.377 |
| | | | 1.513 | 0.988 | 0.990 | 1.545 |
| 09P34 | 1.64 | $^RQ_2(15)$ | 392.385 | 418.509 | 390.949 | 416.978 |
| | | | 1.266 | -1.350 | -1.233 | 1.319 |
| | | | 0.302 | -0.168 | 0.265 | -0.149 |
| | 0.11 | $^RQ_2(16)$ | 418.509 | 444.626 | 416.978 | 442.999 |
| | | | 2.039 | -2.166 | -1.991 | 2.120 |
| | | | 0.111 | 0.109 | 0.098 | 0.102 |
| | -1.52 | $^RQ_2(17)$ | 444.626 | 470.737 | 442.999 | 469.014 |
| | | | 1.487 | -1.574 | -1.456 | 1.544 |
| | | | -0.159 | 0.376 | -0.149 | 0.351 |
| 09P36 | -1.70 | $^PR_2(30)$ | 783.349 | 809.328 | 806.338 | 832.207 |
| | | | 0.543 | -0.561 | -0.560 | 0.578 |
| | | | 0.984 | 1.350 | 1.351 | 0.985 |
| 09P38 | 0.82 | $^RR_1(2)$ | 52.351 | 78.526 | 78.239 | 104.316 |
| | | | 0.087 | -0.139 | -0.086 | 0.148 |
| | | | 0.984 | 1.600 | 2.560 | 0.989 |
| | -0.08 | $^RP_3(13)$ | 340.093 | 366.227 | 312.806 | 338.850 |
| | | | 0.750 | -0.812 | -0.650 | 0.715 |
| | | | 1.725 | 0.988 | 0.993 | 1.800 |
| 09R04 | -1.20 | $^RP_8(11)$ | 287.575 | 313.698 | 260.493 | 286.526 |
| | | | 0.003 | -0.004 | -0.001 | 0.002 |
| | | | 1.656 | 0.991 | 1.336 | 2.321 |
| | 0.12 | $^RP_9(19)$ | 496.276 | 522.341 | 468.487 | 494.464 |
|  |  |  | 0.077 | -0.083 | -0.065 | 0.071 |
|  |  |  | 1.566 | 0.989 | 1.008 | 1.663 |
|  | 0.92 | $^RQ_7(38)$ | 990.136 | 1015.980 | 986.491 | 1012.240 |
|  |  |  | 0.265 | -0.272 | -0.261 | 0.268 |
|  |  |  | 0.726 | 0.731 | 0.713 | 0.728 |
| 09R06 | 1.19 | $^RQ_7(23)$ | 600.767 | 626.808 | 598.571 | 624.517 |
|  |  |  | 0.552 | -0.580 | -0.533 | 0.561 |
|  |  |  | 0.504 | -1.045 | 0.458 | 0.487 |
|  | -2.25 | $^RR_4(24)$ | 627.113 | 653.157 | 650.766 | 676.703 |
|  |  |  | 0.630 | -0.658 | -0.645 | 0.673 |
|  |  |  | 0.978 | 1.975 | 1.094 | 0.984 |
|  | -1.10 | $^RQ_7(24)$ | 626.808 | 652.839 | 624.517 | 650.452 |
|  |  |  | 0.569 | -0.597 | -0.551 | 0.579 |
|  |  |  | 1.065 | 2.192 | -0.496 | -1.038 |
| 09R08 | 1.32 | $^RR_4(26)$ | 679.190 | 705.213 | 702.629 | 728.544 |
|  |  |  | 0.864 | -0.898 | -0.883 | 0.917 |
|  |  |  | 0.985 | 1.414 | 1.428 | 0.985 |
| 09R14 | 1.44 | $^RP_{12}(35)$ | 911.233 | 937.090 | 882.105 | 907.879 |
|  |  |  | 0.013 | -0.014 | -0.012 | 0.013 |
|  |  |  | 1.295 | 0.988 | 0.992 | 1.315 |
| 09R16 | 2.27 | $^RQ_8(27)$ | 704.714 | 730.707 | 702.135 | 728.032 |
|  |  |  | 0.225 | -0.235 | -0.218 | 0.228 |
|  |  |  | 0.573 | -0.474 | 0.549 | -0.447 |
|  | -0.40 | $^RQ_8(28)$ | 730.707 | 756.687 | 728.032 | 753.916 |
|  |  |  | 0.421 | -0.438 | -0.409 | 0.427 |
|  |  |  | 0.264 | 0.606 | 0.249 | 0.587 |
| 09R20 | 1.77 | $^RP_{10}(15)$ | 391.830 | 417.918 | 364.403 | 390.401 |
|  |  |  | 0.003 | -0.004 | -0.002 | 0.003 |
|  |  |  | 1.717 | 0.702 | 1.031 | 2.026 |
|  | 0.28 | $^RR_6(18)$ | 470.514 | 496.605 | 494.791 | 520.778 |
|  |  |  | 1.813 | -1.935 | -1.851 | 1.975 |
|  |  |  | 0.985 | 1.570 | 1.624 | 0.986 |
|  | -0.79 | $^RP_{12}(31)$ | 807.664 | 833.577 | 778.870 | 804.699 |
|  |  |  | 0.019 | -0.020 | -0.017 | 0.018 |
| | | | 1.348 | 0.988 | 0.994 | 1.377 |
| 09R26 | 0.76 | $^RR_5(32)$ | 835.035 | 860.979 | 857.818 | 883.652 |
| | | | 0.600 | -0.619 | -0.610 | 0.630 |
| | | | 0.984 | 1.320 | 1.329 | 0.985 |
| | -0.50 | $^RQ_9(38)$ | 989.670 | 1015.500 | 986.026 | 1011.760 |
| | | | 0.270 | -0.278 | -0.266 | 0.273 |
| | | | 0.727 | 0.732 | 0.712 | 0.728 |
| 09R28 | -0.86 | $^RR_7(15)$ | 392.124 | 418.231 | 416.705 | 442.710 |
| | | | 0.549 | -0.604 | -0.558 | 0.614 |
| | | | 0.985 | 1.623 | 1.731 | 0.986 |
| | 1.37 | $^RQ_9(26)$ | 678.540 | 704.538 | 676.058 | 701.959 |
| | | | 0.436 | -0.456 | -0.420 | 0.441 |
| | | | 0.559 | 0.001 | 0.525 | 0.007 |
| | -1.21 | $^RQ_9(27)$ | 704.538 | 730.524 | 701.959 | 727.849 |
| | | | 0.453 | -0.474 | -0.439 | 0.460 |
| | | | -0.001 | 0.592 | -0.007 | 0.566 |
| 09R30 | -1.67 | $^RR_5(35)$ | 912.826 | 938.728 | 935.276 | 961.065 |
| | | | 0.330 | -0.340 | -0.336 | 0.346 |
| | | | 0.984 | 1.282 | 1.289 | 0.984 |
| 09R32 | 1.41 | $^RR_8(9)$ | 235.314 | 261.446 | 260.493 | 286.526 |
| | | | 0.072 | -0.136 | -0.071 | 0.136 |
| | | | 0.981 | 1.435 | 2.596 | 0.989 |
| | -1.05 | $^RR_6(27)$ | 705.005 | 731.008 | 728.331 | 754.226 |
| | | | 1.260 | -1.310 | -1.283 | 1.333 |
| | | | 0.985 | 1.395 | 1.412 | 0.985 |
| 09R34 | -2.00 | $^RR_5(38)$ | 990.486 | 1016.340 | 1012.600 | 1038.340 |
| | | | 0.209 | -0.215 | -0.212 | 0.218 |
| | | | 0.988 | 1.253 | 1.261 | 0.984 |
| | -0.73 | $^RP_{14}(38)$ | 988.007 | 1013.800 | 958.647 | 984.356 |
| | | | 0.002 | -0.002 | -0.002 | 0.002 |
| | | | 55.899 | 147.422 | 0.660 | 15.354 |
| 10P04 | 0.76 | $^PR_{12}(20)$ | 521.859 | 547.891 | 545.805 | 571.729 |
| | | | 0.012 | -0.013 | -0.014 | 0.015 |
| | | | 0.984 | 1.567 | 1.491 | 0.987 |
|       | 1.29  | $^PR_{13}(28)$ | 729.583 | 755.524 | 752.621 | 778.449 |
|       |       |             | 0.005   | -0.005  | -0.005  | 0.005   |
|       |       |             | 0.967   | 1.516   | 1.450   | -39.150 |
|       | 1.61  | $^PQ_9(30)$ | 782.461 | 808.410 | 779.513 | 805.364 |
|       |       |             | 0.354   | -0.368  | -0.360  | 0.374   |
|       |       |             | 0.623   | 0.008   | 0.620   | 0.016   |
|       | -1.43 | $^PQ_9(31)$ | 808.410 | 834.347 | 805.364 | 831.201 |
|       |       |             | 0.361   | -0.375  | -0.367  | 0.380   |
|       |       |             | -0.008  | 0.651   | -0.016  | 0.661   |
| 10P06 | 0.79  | $^PP_7(18)$ | 470.424 | 496.510 | 442.677 | 468.672 |
|       |       |             | 0.557   | -0.598  | -0.549  | 0.588   |
|       |       |             | 1.656   | 1.005   | 0.980   | 1.585   |
|       | 1.20  | $^PR_{12}(18)$ | 469.769 | 495.818 | 493.931 | 519.872 |
|       |       |             | 0.007   | -0.007  | -0.008  | 0.009   |
|       |       |             | 150.647 | 73.864  | 32.298  | 0.564   |
|       | 2.02  | $^PP_6(25)$ | 652.964 | 678.990 | 624.597 | 650.535 |
|       |       |             | 0.857   | -0.894  | -0.835  | 0.872   |
|       |       |             | 1.462   | 0.988   | 0.986   | 1.445   |
|       | 0.84  | $^PP_5(32)$ | 835.035 | 860.979 | 806.069 | 831.928 |
|       |       |             | 0.545   | -0.563  | -0.531  | 0.548   |
|       |       |             | 1.338   | 0.988   | 0.987   | 1.334   |
|       | 0.21  | $^PR_{14}(34)$ | 884.690 | 910.542 | 907.007 | 932.741 |
|       |       |             | 0.003   | -0.003  | -0.003  | 0.003   |
|       |       |             | -0.300  | 1.457   | 1.481   | 0.928   |
| 10P08 | 1.93  | $^PR_{12}(16)$ | 417.648 | 443.712 | 442.024 | 467.981 |
|       |       |             | 0.002   | -0.003  | -0.003  | 0.004   |
|       |       |             | 0.982   | 1.728   | 1.545   | 0.988   |
| 10P12 | 0.93  | $^PQ_{10}(32)$ | 834.114 | 860.030 | 830.955 | 856.771 |
|       |       |             | 0.125   | -0.129  | -0.127  | 0.131   |
|       |       |             | 0.654   | 0.665   | 0.661   | 0.678   |
| 10P14 | 2.06  | $^PP_6(32)$ | 834.899 | 860.840 | 805.929 | 831.785 |
|       |       |             | 0.555   | -0.574  | -0.542  | 0.560   |
|       |       |             | 1.338   | 0.988   | 0.987   | 1.332   |
| 10P16 | 1.39  | $^PP_9(13)$ | 339.732 | 365.839 | 312.445 | 338.459 |
| | | | 0.297 | -0.359 | -0.325 | 0.381 |
| | | | 1.926 | 0.988 | 0.936 | 1.543 |
| | 2.03 | $^PP_8(20)$ | 522.472 | 548.535 | 494.544 | 520.518 |
| | | | 0.263 | -0.281 | -0.260 | 0.277 |
| | | | 1.589 | 0.988 | 0.980 | 1.528 |
| | 1.51 | $^PR_{14}(23)$ | 599.475 | 625.460 | 623.046 | 648.920 |
| | | | 0.001 | -0.001 | -0.001 | 0.002 |
| | | | 0.721 | 1.810 | 1.702 | -0.363 |
| | 0.17 | $^PP_7(27)$ | 704.870 | 730.868 | 676.326 | 702.238 |
| | | | 0.531 | -0.553 | -0.519 | 0.541 |
| | | | 1.422 | 0.988 | 0.986 | 1.406 |
| 10P18 | -1.52 | $^PR_{13}(13)$ | - | 365.368 | - | 389.938 |
| | | | - | 0.000 | - | 0.000 |
| | | | - | 19721.500 | - | 13439.500 |
| | -0.65 | $^PP_9(15)$ | 391.940 | 418.034 | 364.467 | 390.470 |
| | | | 0.487 | -0.554 | -0.506 | 0.567 |
| | | | 1.790 | 0.988 | 0.959 | 1.571 |
| | 1.81 | $^PQ_{11}(18)$ | 469.927 | 495.985 | 468.148 | 494.107 |
| | | | 0.041 | -0.046 | -0.045 | 0.050 |
| | | | 0.304 | -0.229 | 0.447 | -0.305 |
| | -0.06 | $^PQ_{11}(19)$ | 495.985 | 522.035 | 494.107 | 520.057 |
| | | | 0.077 | -0.085 | -0.083 | 0.091 |
| | | | 0.138 | 0.194 | 0.183 | 0.190 |
| | -2.04 | $^PQ_{11}(20)$ | 522.035 | 548.076 | 520.057 | 546.000 |
| | | | 0.047 | -0.052 | -0.051 | 0.055 |
| | | | -0.347 | 4.871 | -0.340 | -5.539 |
| | -1.27 | $^PP_8(22)$ | 574.589 | 600.634 | 546.483 | 572.440 |
| | | | 0.357 | -0.377 | -0.351 | 0.371 |
| | | | 1.534 | 0.988 | 0.932 | 1.491 |
| 10P20 | -1.14 | $^PQ_{11}(38)$ | 989.090 | 1014.910 | 985.311 | 1011.030 |
| | | | 0.043 | -0.045 | -0.044 | 0.045 |
| | | | 0.733 | 0.739 | 0.734 | 0.749 |
| 10P24 | 1.08 | $^PR_{14}(14)$ | - | 391.281 | - | 415.725 |
| | | | - | 0.000 | - | 0.000 |
| | | | - | 0.976 | - | 0.997 |
| 10P26 | -1.93 | $^PP_{10}(16)$ | 417.918 | 443.998 | 390.355 | 416.344 |
| | | | 0.102 | -0.116 | -0.107 | 0.120 |
| | | | 1.761 | 0.988 | 0.956 | 1.539 |
| | 1.71 | $^PR_{16}(27)$ | 702.741 | 728.660 | 725.802 | 751.607 |
| | | | 0.000 | 0.000 | 0.000 | 0.000 |
| | | | 0.975 | 1.409 | 1.366 | 0.940 |
| | -0.53 | $^PQ_{12}(31)$ | 807.664 | 833.577 | 804.576 | 830.387 |
| | | | 0.099 | -0.103 | -0.101 | 0.105 |
| | | | 0.641 | 0.653 | 0.654 | 0.671 |
| 10P28 | -0.83 | $^PP_{10}(18)$ | 470.072 | 496.137 | 442.326 | 468.301 |
| | | | 0.176 | -0.194 | -0.179 | 0.196 |
| | | | 1.670 | 0.988 | 0.968 | 1.530 |
| 10P30 | 0.64 | $^PP_{10}(20)$ | 522.195 | 548.245 | 494.268 | 520.227 |
| | | | 0.195 | -0.211 | -0.196 | 0.211 |
| | | | 1.599 | 0.988 | 0.974 | 1.505 |
| | -1.00 | $^PR_{16}(22)$ | 572.980 | 598.952 | 596.608 | 622.468 |
| | | | 0.000 | 0.000 | 0.000 | 0.000 |
| | | | -6.144 | 2.438 | 2.368 | 0.744 |
| 10P32 | 1.46 | $^PQ_{13}(25)$ | 651.692 | 677.667 | 649.192 | 675.066 |
| | | | 0.021 | -0.022 | -0.022 | 0.023 |
| | | | 0.523 | -0.047 | 0.568 | -0.043 |
| | -1.14 | $^PQ_{13}(26)$ | 677.667 | 703.631 | 675.066 | 700.929 |
| | | | 0.024 | -0.025 | -0.025 | 0.026 |
| | | | 0.043 | 0.567 | 0.040 | 0.600 |
| 10P36 | -1.95 | $^PP_9(33)$ | 860.270 | 886.179 | 831.201 | 857.025 |
| | | | 0.213 | -0.220 | -0.209 | 0.216 |
| | | | 1.321 | 0.988 | 0.986 | 1.310 |
| 10P38 | 0.04 | $^PQ_{14}(23)$ | 599.475 | 625.460 | 597.162 | 623.046 |
| | | | 0.013 | -0.014 | -0.014 | 0.015 |
| | | | 0.472 | 0.496 | 0.541 | 0.559 |
| | -0.12 | $^PP_9(35)$ | 912.075 | 937.955 | 882.834 | 908.629 |
| | | | 0.305 | -0.315 | -0.300 | 0.309 |
| | | | 1.295 | 0.988 | 0.986 | 1.286 |
| | -0.63 | $^PR_{19}(35)$ | 908.370 | 934.145 | 930.358 | 956.010 |
| | | | 0.000 | 0.000 | 0.000 | 0.000 |
| | | | 13483.300 | 2600.890 | 602.381 | -112.979 |
| 10P40 | -1.78 | $^PR_{17}(17)$ | - | 468.779 | - | 492.823 |
| | | | - | 0.000 | - | 0.000 |
| | | | - | 1.026 | - | 0.975 |
| | 2.13 | $^PP_9(37)$ | 963.820 | 989.670 | 934.410 | 960.175 |
| | | | 0.132 | -0.136 | -0.130 | 0.134 |
| | | | 1.270 | 0.988 | 0.987 | 1.263 |
| 10R04 | 1.32 | $^PR_{10}(13)$ | 339.637 | 365.737 | 364.360 | 390.355 |
| | | | 0.003 | -0.004 | -0.005 | 0.005 |
| | | | 0.980 | 1.858 | 1.601 | 0.987 |
| | 1.11 | $^PP_5(23)$ | 600.979 | 627.030 | 572.788 | 598.751 |
| | | | 0.800 | -0.838 | -0.776 | 0.813 |
| | | | 1.505 | 0.989 | 0.986 | 1.489 |
| | 0.86 | $^PQ_8(25)$ | 652.695 | 678.710 | 650.251 | 676.168 |
| | | | 0.419 | -0.440 | -0.429 | 0.449 |
| | | | 0.515 | 0.166 | 0.557 | 0.174 |
| | -1.69 | $^PQ_8(26)$ | 678.710 | 704.714 | 676.168 | 702.073 |
| | | | 0.305 | -0.318 | -0.311 | 0.324 |
| | | | -0.240 | 0.566 | -0.251 | 0.580 |
| 10R08 | 0.97 | $^PP_4(27)$ | 705.213 | 731.224 | 676.676 | 702.601 |
| | | | 0.837 | -0.869 | -0.811 | 0.842 |
| | | | 1.421 | 0.988 | 0.987 | 1.416 |
| | 2.20 | $^PP_3(34)$ | 887.119 | 913.042 | 857.992 | 883.831 |
| | | | 0.713 | -0.734 | -0.692 | 0.713 |
| | | | 1.311 | 0.988 | 0.988 | 1.311 |
| 10R12 | -1.80 | $^PR_{10}(20)$ | 522.195 | 548.245 | 546.178 | 572.119 |
| | | | 0.019 | -0.020 | -0.021 | 0.023 |
| | | | 0.975 | 1.591 | 1.532 | 0.825 |
| | 1.18 | $^PP_4(24)$ | 627.113 | 653.157 | 598.836 | 624.793 |
| | | | 0.876 | -0.913 | -0.847 | 0.884 |
| | | | 1.482 | 0.988 | 0.987 | 1.473 |
| | 0.87 | $^PQ_7(33)$ | 860.675 | 886.597 | 857.452 | 883.275 |
| | | | 0.421 | -0.435 | -0.425 | 0.438 |
| | | | 0.666 | 0.676 | 0.668 | 0.684 |
| 10R14 | 0.09 | $^PP_6(8)$ | 209.263 | 235.411 | 182.434 | 208.487 |
| | | | 0.487 | -0.690 | -0.591 | 0.761 |
| | | | 2.304 | 0.989 | 0.893 | 1.506 |
| | 0.84 | $^PQ_7(16)$ | 418.231 | 444.331 | 416.674 | 442.677 |
| | | | 0.454 | -0.494 | -0.481 | 0.519 |
| | | | 0.278 | -0.006 | 0.356 | -0.003 |
| | -0.82 | $^PQ_7(17)$ | 444.331 | 470.424 | 442.677 | 468.672 |
| | | | 0.501 | -0.541 | -0.526 | 0.564 |
| | | | 0.006 | 0.344 | 0.003 | 0.397 |
| | -0.38 | $^PR_{11}(30)$ | 782.002 | 807.936 | 804.863 | 830.684 |
| | | | 0.024 | -0.025 | -0.025 | 0.026 |
| | | | 0.984 | 1.345 | 1.330 | 0.985 |
| 10R18 | 0.00 | $^PP_2(34)$ | 887.185 | 913.109 | 858.062 | 883.903 |
| | | | 0.731 | -0.752 | -0.708 | 0.729 |
| | | | 1.311 | 0.988 | 0.988 | 1.312 |
| 10R20 | -1.09 | $^PP_5(11)$ | 287.740 | 313.879 | 260.629 | 286.675 |
| | | | 0.526 | -0.595 | -0.521 | 0.584 |
| | | | 1.888 | 0.988 | 0.973 | 1.715 |
| | 1.65 | $^PR_{10}(27)$ | 704.341 | 730.320 | 727.560 | 753.428 |
| | | | 0.030 | -0.031 | -0.032 | 0.033 |
| | | | 0.984 | 1.398 | 1.379 | 0.985 |
| 10R22 | -0.83 | $^PQ_6(33)$ | 860.840 | 886.766 | 857.627 | 883.455 |
| | | | 1.190 | -1.228 | -1.198 | 1.236 |
| | | | 0.665 | 0.675 | 0.666 | 0.683 |
| | -1.96 | $^PR_{11}(37)$ | 963.255 | 989.090 | 985.311 | 1011.030 |
| | | | 0.010 | -0.010 | -0.010 | 0.011 |
| | | | 0.978 | 1.253 | 1.248 | 0.983 |
| 10R24 | 2.11 | $^PR_8(14)$ | 365.930 | 392.038 | 390.573 | 416.577 |
| | | | 0.022 | -0.025 | -0.027 | 0.030 |
| | | | 0.980 | 1.764 | 1.643 | 0.937 |
| | 1.10 | $^PQ_6(17)$ | 444.416 | 470.514 | 442.767 | 468.767 |
| | | | 1.379 | -1.480 | -1.434 | 1.531 |

|  |  |  | 0.318 | -0.049 | 0.369 | -0.050 |
|---|---|---|---|---|---|---|
|  | -0.64 | $^PQ_6(18)$ | 470.514 | 496.605 | 468.767 | 494.760 |
|  |  |  | 1.677 | -1.790 | -1.735 | 1.844 |
|  |  |  | 0.043 | 0.377 | 0.044 | 0.413 |
|  | 0.93 | $^PR_9(22)$ | 574.446 | 600.484 | 598.228 | 624.159 |
|  |  |  | 0.121 | -0.129 | -0.133 | 0.140 |
|  |  |  | 0.984 | 1.506 | 1.473 | 0.986 |
|  | 0.56 | $^PP_2(30)$ | 783.349 | 809.328 | 754.562 | 780.456 |
|  |  |  | 0.935 | -0.965 | -0.903 | 0.934 |
|  |  |  | 1.369 | 0.988 | 0.988 | 1.370 |
| 10R28 | 0.92 | $^PR_7(9)$ | 235.366 | 261.504 | 260.531 | 286.568 |
|  |  |  | 0.008 | -0.011 | -0.014 | 0.017 |
|  |  |  | 0.979 | 2.040 | 1.663 | 0.989 |
|  | 2.23 | $^PR_8(17)$ | 444.233 | 470.320 | 468.563 | 494.544 |
|  |  |  | 0.039 | -0.043 | -0.045 | 0.048 |
|  |  |  | 0.984 | 1.644 | 1.579 | 0.987 |
|  | -1.55 | $^PR_9(25)$ | 652.532 | 678.540 | 675.989 | 701.888 |
|  |  |  | 0.109 | -0.115 | -0.118 | 0.123 |
|  |  |  | 0.984 | 1.438 | 1.417 | 0.986 |
| 10R30 | -1.18 | $^PP_3(19)$ | 496.803 | 522.895 | 468.967 | 494.971 |
|  |  |  | 2.037 | -2.147 | -1.950 | 2.058 |
|  |  |  | 1.600 | 0.988 | 0.987 | 1.588 |
|  | -1.47 | $^PR_{10}(35)$ | 911.820 | 937.693 | 934.133 | 959.891 |
|  |  |  | 0.025 | -0.025 | -0.026 | 0.026 |
|  |  |  | 0.984 | 1.276 | 1.270 | 0.984 |
| 10R32 | 0.35 | $^PP_2(25)$ | 653.273 | 679.311 | 624.913 | 650.865 |
|  |  |  | 1.234 | -1.283 | -1.186 | 1.235 |
|  |  |  | 1.459 | 0.988 | 0.988 | 1.459 |
|  | -1.96 | $^PR_9(28)$ | 730.524 | 756.498 | 753.650 | 779.513 |
|  |  |  | 0.093 | -0.097 | -0.098 | 0.102 |
|  |  |  | 0.973 | 1.383 | 1.370 | 0.984 |
|  | 0.60 | $^PQ_5(38)$ | 990.486 | 1016.340 | 986.788 | 1012.540 |
|  |  |  | 0.517 | -0.531 | -0.519 | 0.532 |
|  |  |  | 0.727 | 0.733 | 0.723 | 0.738 |
| 10R34 | 1.21 | $^PP_4(9)$ | 235.480 | 261.631 | 208.553 | 234.611 |
| | | | 0.447 | -0.517 | -0.438 | 0.502 |
| | | | 1.957 | 0.988 | 0.973 | 1.755 |
| | 1.96 | $^PQ_5(27)$ | 705.119 | 731.126 | 702.501 | 728.411 |
| | | | 0.720 | -0.748 | -0.726 | 0.754 |
| | | | 0.564 | -0.262 | 0.573 | -0.258 |
| | -0.76 | $^PQ_5(28)$ | 731.126 | 757.122 | 728.411 | 754.310 |
| | | | 1.079 | -1.119 | -1.088 | 1.128 |
| | | | 0.182 | 0.604 | 0.179 | 0.604 |
| | 1.27 | $^PP_1(31)$ | 809.364 | 835.331 | 780.496 | 806.379 |
| | | | 0.709 | -0.731 | -0.684 | 0.706 |
| | | | 1.353 | 0.988 | 0.991 | 1.356 |
| 10R36 | 1.95 | $^PQ_5(8)$ | 209.297 | 235.449 | 208.523 | 234.578 |
| | | | 0.136 | -0.172 | -0.166 | 0.199 |
| | | | -0.101 | 0.008 | 0.270 | -0.158 |
| | 1.08 | $^PQ_5(9)$ | 235.449 | 261.596 | 234.578 | 260.629 |
| | | | 0.282 | -0.338 | -0.326 | 0.377 |
| | | | -0.005 | -0.020 | 0.097 | -0.087 |
| | 0.11 | $^PQ_5(10)$ | 261.596 | 287.740 | 260.629 | 286.675 |
| | | | 0.460 | -0.532 | -0.513 | 0.580 |
| | | | 0.014 | 0.007 | 0.064 | 0.019 |
| | -0.96 | $^PQ_5(11)$ | 287.740 | 313.879 | 286.675 | 312.717 |
| | | | 0.492 | -0.557 | -0.536 | 0.597 |
| | | | -0.008 | 0.048 | -0.021 | 0.086 |
| | -2.12 | $^PQ_5(12)$ | 313.879 | 340.012 | 312.717 | 338.754 |
| | | | 0.373 | -0.416 | -0.400 | 0.440 |
| | | | -0.072 | 0.189 | -0.128 | 0.276 |
| | -0.29 | $^PR_9(31)$ | 808.410 | 834.347 | 831.201 | 857.025 |
| | | | 0.147 | -0.152 | -0.154 | 0.160 |
| | | | 0.984 | 1.332 | 1.322 | 0.984 |
| | 1.57 | $^RP_0(37)$ | 964.971 | 990.852 | 935.606 | 961.404 |
| | | | 0.368 | -0.378 | -0.357 | 0.366 |
| | | | 1.242 | 0.962 | 0.962 | 1.244 |
| | 1.70 | $^PP_0(37)$ | 964.971 | 990.852 | 935.609 | 961.408 |
|       |       |             | 0.349   | -0.358  | -0.338  | 0.347   |
|       |       |             | 1.311   | 1.016   | 1.015   | 1.314   |
| 10R38 | 0.05  | $^PR_8(24)$ | 626.670 | 652.695 | 650.251 | 676.168 |
|       |       |             | 0.133   | -0.140  | -0.143  | 0.149   |
|       |       |             | 0.984   | 1.459   | 1.438   | 0.986   |
|       | -1.41 | $^RP_0(36)$ | 939.075 | 964.971 | 909.792 | 935.606 |
|       |       |             | 0.434   | -0.445  | -0.420  | 0.432   |
|       |       |             | 1.319   | 1.013   | 1.012   | 1.321   |
|       | -1.29 | $^PP_0(36)$ | 939.075 | 964.971 | 909.796 | 935.609 |
|       |       |             | 0.455   | -0.467  | -0.441  | 0.453   |
|       |       |             | 1.256   | 0.965   | 0.964   | 1.259   |